\documentclass[aps,floats, floatfix, nofootinbib]{revtex4}

\usepackage{graphicx}
\usepackage{amsmath}
\usepackage{amssymb}
\usepackage{color}

\def\ltwid{\mathrel{\raise.3ex\hbox{$<$\kern-.75em\lower1ex\hbox{$\sim$}}}}
\def\gtwid{\mathrel{\raise.3ex\hbox{$>$\kern-.75em\lower1ex\hbox{$\sim$}}}}

\def\overleftrightarrow#1{\vbox{\ialign{##\crcr
     $\leftrightarrow$\crcr\noalign{\kern-1pt\nointerlineskip}
     $\hfil\displaystyle{#1}\hfil$\crcr}}}

\def\zetaa{\zeta_{1}}
\def\zetaaa{\zeta_{2}}

\newcommand{\be}{\begin{equation}}
\newcommand{\ee}{\end{equation}}
\newcommand{\bea}{\begin{eqnarray}}
\newcommand{\eea}{\end{eqnarray}}
\newcommand{\nn}{\nonumber}

\newcommand{\ec}[1]{Eq.~(\ref{eq:#1})}
\newcommand{\Ec}[1]{(\ref{eq:#1})}

\newcommand{\eql}[1]{\label{eq:#1}}

\newcommand\osq{\overline{\square}}
\newcommand\ovr{\overline{R}}

\newcommand\ovX{\overline{X}}
\newcommand\ovf{\overline{f}}
\newcommand\ovU{\overline{U}}

\begin{document}

\title{Revival of the Deser-Woodard nonlocal gravity model:
\\Comparison of the original nonlocal form and a localized formulation
}

\author{Sohyun Park}
\email{park@fzu.cz}
\affiliation{CEICO, Institute of Physics of the Czech Academy of Sciences, Na Slovance 2, 18221 Prague 8 Czech Republic}

\begin{abstract}
\noindent
We examine the origin of two opposite results for the growth of perturbations in the Deser-Woodard (DW) nonlocal gravity model. 
One group previously analyzed the model in its original nonlocal form and showed that the growth of structure in the DW model is enhanced compared to general relativity (GR) and thus concluded that the model was ruled out.
Recently, however, another group has reanalyzed it by localizing the model and found that the growth in their localized version is suppressed even compared to the one in GR. 
The question was whether the discrepancy originates from an intrinsic difference between the nonlocal and localized formulations or is due to their different implementations of the sub-horizon limit. 
We show that the nonlocal and local formulations give the same solutions for the linear perturbations as long as the initial conditions are set the same. The different implementations of the sub-horizon limit lead to different transient behaviors of some perturbation variables; however, they do not affect the growth of matter perturbations at the sub-horizon scale much. 
In the meantime, we also report an error in the numerical calculation code of the former group and verify that after fixing the error the nonlocal version also gives the suppressed growth. 
Finally, we discuss two alternative definitions of the effective gravitational constant taken by the two groups and some open problems.
\end{abstract}

\maketitle

\section{Introduction}\label{intro}

Nonlocal modifications of gravity  as an account for the current phase of cosmic acceleration 
have recently received significant attention \cite{DW-2007, Joukovskaya-2007,NO-2007,CMN-2008,NO-2008,Koivisto-0803,ghost-nonlocal-Koivisto-0807,CENO-2009,Koshelev-2009,DW-2009,Biswas-2010,ghost-nonlocal-NOSZ-1010,Bamba-1104,Barvinsky-1107,ghost-nonlocal-ZS-1108,EPV-1110,Barvinsky-1112,PD-2012,Barvinsky-1209,EPV-1209,EPVZ-1302,Maggiore-1307,DW-2013,DP-2013,Maggiore-1311.3421,Maggiore-1311.3435,W-review-2014,Maggiore-1401,MM-2014,Maggiore-1403,Koivisto-1406,BLHBP-1408,DM-1408,Maggiore-1411,ZWLLCCS-1511,Maggiore-1512,ZKSZ-1601,Maggiore-1602.01078,Maggiore-1602.03558,Maggiore-1603,NAAKR-1606,Maggiore-review,PS-2016, WZWZC-1611,NCA-2017,DDSKR-1701,VAAS-2017,Dirian-1704,ABN-1707,Roobiat-2017,ZWYZCZ-2017,Sebastian-1708,Sebastian-1709,Bamba-1711}.
\footnote{For nonlocal gravity with more quantum field theory oriented approaches, see Refs. \cite{Parker-1985, Banks-1988, Wetterich-1998, Barvinsky-2003, Hamber-2005, Biswas-2006, Mazzitelli-2007, Ferreira-2013, Donoghue-2014, Calmet-2015, Donoghue-2015, Oda-1704, Oda-1706, Oda-1709,Mazzitelli-2017, Bautista-2017, Calmet-2017,Oda-1711}; as an account for inflation, see Refs.
\cite{NO-2007, TW-0904, TW-1006, TW-1405, KKM-1604, TW-1606}; 
and for dark matter, see Refs. \cite{SW-2003,DEW-1106,W-MOND-review-1403,DEW-1405, KRSTWX-1608}. 
}  
One such attempt is a model proposed by Deser and Woodard, in which a nonlocal piece in the form of $Rf(\square^{-1} R) $ is added to the Einstein-Hilbert term \cite{DW-2007}. Here, $f$ is a free function with the nonlocal argument, 
the inverse scalar d'Alembertian acting on the Ricci scalar. 

An advantage of leaving $f$ as an arbitrary function 
is that $f$ can be constructed to reproduce any desirable expansion history. 
A generic technique for how to tune the function $f$ for any given expansion history 
has been provided in Ref. \cite{DW-2009}. \footnote{Similar reconstruction techniques are found in Refs. \cite{EPV-1209,EPVZ-1302}.}
Applying this technique to the special case of Lambda-cold-dark-matter ($\Lambda$CDM) cosmology, the authors of \cite{DW-2009} have obtained a numerical solution for $f$ that exactly reproduces the expansion history of $\Lambda$CDM without a cosmological constant.
This implies that one does not test the Deser-Woodard (DW) model with background expansion data. 
One rather fixes the model by adjusting the function $f$ such that its background is identical to the given expansion and then examines how perturbations grow. 
A key aspect is that, even though its background expansion is the same as in GR, the growth of perturbations in the DW model is different from the growth in general relativity (GR). 
Whether one fixes its background to the $\Lambda$CDM expansion history as in Refs. \cite{PD-2012, DP-2013, NCA-2017} or to some non-$\Lambda$CDM as in Ref. \cite{PS-2016}, one gets different structure formation from that in GR.
Thus, the model is testable with growth data (but not with expansion data).    

This is a distinct feature of the DW nonlocal model. 
Other classes of nonlocal models, for instance, the $m^2 R \square^{-2} R$ model proposed by Maggiore and Mancarella \cite{MM-2014} or $m^2 \square^{-1} R$ model by Vardanyan \textit{et al.} 
\cite{VAAS-2017},  
do not reproduce a chosen background expansion. 
Thus, one tests these nonlocal models with both expansion and growth data.
However, in practice, one first fixes the value (or range) of the parameter $m$ in such a way that the model describes the expansion data as closely as possible and then investigates how perturbations evolve.
Therefore, when it comes to testing the growth of perturbations, no free parameter (or free function in the case of the DW model) is left.  The models then make their unique predictions for the growth, which distinguish them from each other.    

In this regard, measuring the growth of structure is a powerful test that can decisively rule out a gravity model \cite{Bertschinger-2006,Hu-Sawicki-0708,Baker:2012zs,Huterer:2013xky}. 
For the DW nonlocal gravity model, two groups (one in its original nonlocal form  \cite{PD-2012, DP-2013} and the other in a localized version \cite{NCA-2017}) have studied its prediction for the growth once the background expansion was fixed to the one of $\Lambda$CDM, and 
they have obtained results opposite of one another.
The reason for this discrepancy has been suspected whether it is due to an intrinsic difference between the nonlocal and localized formulations, or it is due to their different implementations of the sub-horizon limit.
In the present paper, we thoroughly examine the two formulations and the different approximations that they have taken for the sub-horizon scale. 

\section{Model and Zeroth-Order Equations}  
\label{sec-zeroth-order}

The DW model nonlocally modifies the Einstein-Hibert action as \cite{DW-2007}
\be
 S_{DW} = \frac{1}{16\pi G} \int d^4 x \sqrt{-g} R \bigg[1 + f\Big(\frac{1}{\square}R\Big)\bigg] \;,
 \eql{action}
\ee
where $f$ is a function of the inverse d'Alembertian acting on the Ricci scalar that will be determined by matching a given expansion history. 
Note that the argument $\square^{-1}R$ is dimensionless, thus no new mass scale is introduced in the action.  
By varying the action with respect to the metric and imposing the retarded boundary conditions, \footnote{The retarded boundary condition on Green's function ensures the field equations are causal. Note also that 
Green's function being the inverse of a differential operator, is sufficient for conservation. 
Hence, the resulting field equations are causal and conserved. See Ref. \cite{W-review-2014} for more discussions on the causality and conservation.}  
one obtains the modified Einstein equations
\be
G_{\mu\nu} + \Delta G_{\mu\nu} = 8\pi G T_{\mu\nu} \;,
\eql{nonlocal_field_eq}
\ee
where the correction to the Einstein tensor is \cite{DW-2007}
\be
\Delta G_{\mu\nu} =
\Bigl[ G_{\mu\nu} \!+\! g_{\mu\nu}\square \!-\! D_{\mu}D_{\nu} \Bigr]
\biggl\{\! f\Big(\frac{1}{\square}R\Big)  \!+\! \frac{1}{\square}\Bigl[R f'
\Big(\frac{1}{\square}R\Big)\Bigr] \!\biggr\}
 +
\Bigl[ \delta_{\mu}^{(\rho}\delta_{\nu}^{\sigma)} 
\!-\! \frac{1}{2}g_{\mu\nu}g^{\rho\sigma}\Bigl] 
\partial_{\rho} \Big(\frac{1}{\square}R\Big)
\partial_{\sigma}
\biggl(\frac{1}{\square}\Bigl[R f'
\Big(\frac{1}{\square}R\Big)\Bigr]\biggr) \;.
\eql{DeltaGmn}
\ee
Here, a prime means a derivative of the function with respect to the argument.
Applying the field equations \Ec{nonlocal_field_eq} to the Friedman-Lema\^{i}tre-Robertson-Walker (FLRW) metric,
\be
ds^2 = -dt^2 +a^2(t) d\vec{x}\cdot d\vec{x} \;,
\ee
leads to
\bea
3H^2 + \Delta \overline{G}_{00} &=& 8\pi G \rho \;, \\
-2\dot{H} -3H^2 + \frac{1}{3a^2}\delta^{ij}\Delta \overline{G}_{ij} &=& 8\pi G p \;.
\eea
Here, $\rho$ and $p$ are the energy density and pressure of a perfect fluid and 
we denote quantities in the FLRW background with overbars to distinguish them from perturbed quantities.
The zeroth-order corrections are \cite{DW-2007}
\bea
\Delta \overline{G}_{00} &\!\!=\!\!& 
\Bigl[ 3H^2 + 3H\partial_t \Big]
\biggl\{f \Bigr(\frac{1}{\osq}\ovr \Bigl) + \frac{1}{\osq}\Bigl[\ovr f'\Bigr(\frac{1}{\osq}\ovr\Bigl)\Bigr] \biggr\}
+\frac{1}{2}
\partial_{t}\Bigr(\frac{1}{\osq}\ovr\Bigl)
\partial_{t}\bigg(\frac{1}{\osq}\Bigl[\overline{R} f'\Bigr(\frac{1}{\osq}\ovr\Bigl)\Bigr]\bigg)\;, 
\eql{DeltaG00_zeroth}\\
\Delta  \overline{G}_{ij} &\!\!=\!\!& 
a^2\delta_{ij}\Bigg[ 
\frac{1}{2}
\partial_{t}\Bigr(\frac{1}{\osq}\ovr\Bigl)
\partial_{t}\bigg(\frac{1}{\overline{\square}}\Bigl[\overline{R} f'\Bigr(\frac{1}{\osq}\ovr\Bigl)\Bigr]\bigg)\;
- \Big( 2\dot{H}+ 3H^2+2H\partial_t + \partial^2_t\Big)
\biggl\{ f \Bigr(\frac{1}{\osq}\ovr \Bigl) + \frac{1}{\osq}\Bigl[\ovr f'\Bigr(\frac{1}{\osq}\ovr\Bigl)\Bigr]\biggr\}
\Bigg]\;.
\eql{DeltaGij_zeroth}
\eea

\subsection{Localization}

One can localize the nonlocal field equations by introducing two auxiliary variables $X$ and $U$ defined as, following the notation of Ref. \cite{NCA-2017}, 
\bea
\Box X &\equiv & R \;, 
\eql{eq:boxx}\\
\Box U &\equiv & R f'(X)  \;. 
\eql{eq:boxu}
\eea
Also, we denote the operator acting on the nonlocal function as
\be
\mathcal{D}_{\mu\nu} \equiv G_{\mu\nu} + g_{\mu\nu}\square - D_{\mu}D_{\nu}\;.
\eql{mathcalD}
\ee
In this notation, the modified Einstein tensor $\Delta G_{\mu\nu}$ is simply written as
\be
\Delta G_{\mu\nu} =\mathcal{D}_{\mu\nu}( f + U ) +
\Bigl[ \delta_{\mu}^{(\rho}\delta_{\nu}^{\sigma)} \!-\! \frac{1}{2}g_{\mu\nu}g^{\rho\sigma}\Bigl] 
\partial_{\rho} X
\partial_{\sigma} U\;,
\eql{DeltaGmn-local}
\ee
and the zeroth-order corrections \Ec{DeltaG00_zeroth} and \Ec{DeltaGij_zeroth} 
become
\bea
\Delta \overline{G}_{00} &\!\!=\!\!& 
[ 3H^2 + 3H\partial_t] (\ovf+\ovU) +\frac{1}{2}\partial_t{\ovX}\partial_t{\ovU}\;,
\eql{DeltaG00_zeroth-local} \\
\Delta  \overline{G}_{ij} &\!\!=\!\!& 
a^2\delta_{ij}\Big[ -( 2\dot{H}+ 3H^2+2H\partial_t + \partial^2_t)(\ovf+\ovU)+\frac{1}{2}\partial_t{\ovX}\partial_t{\ovU}\; 
\Big]\;.
\eql{DeltaGij_zeroth-local}
\eea
We then have the zeroth-order modified Einstein equations 
\bea
3H^2 + [3H^2 + 3H\partial_{t}](\ovf +\ovU) +\frac{1}{2}\partial_t{\ovX}\partial_t{\ovU}
&=& 8\pi G \rho \;, 
\eql{zeroth-00-local-eq}
\\
-(2\dot{H} + 3H^2) -[2\dot{H} + 3H^2 +2H\partial_{t}+\partial_{t}^2](\ovf +\ovU)
+\frac{1}{2}\partial_t{\ovX}\partial_t{\ovU}
&=& 8\pi G p \;.
\eql{zeroth-trace-local-eq}
\eea
accompanied by the zeroth-order auxiliary field equations 
\bea
-(\partial_t^2 + 3H\partial_t)\ovX&=& 6(\dot{H} +2H^2) \;, 
\eql{eq:boxx-local}
\\
-(\partial_t^2 + 3H\partial_t)\ovU&=& 6(\dot{H} +2H^2) \ovf'
\;,   
\eql{eq:boxu-local}
\eea
in this localized formulation.

\subsection{Reconstruction of $f$}

\begin{figure}[htbp]
\centering
\includegraphics[width=12cm]{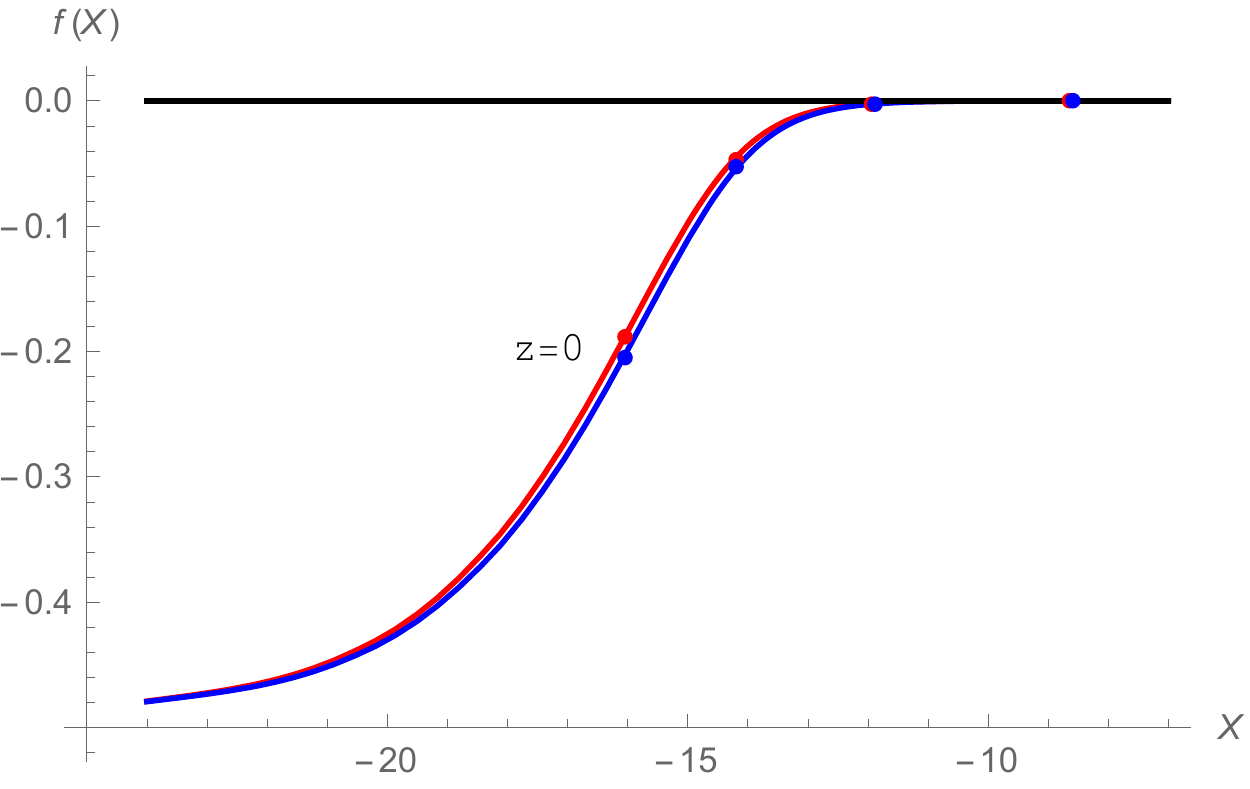}
\caption{The blue curve is $f_{\rm WMAP5year}(\ovX)$ with $\{\Omega_{\Lambda}, \Omega_{m},\Omega_{r}\} = \{0.72, 0.28, 8.5 \times 10^{-5}\}$, which is identical to the plot given in  
Ref. \cite{DW-2009}. The red curve is $f_{\rm Planck 2015}(\ovX)$ with 
$\{\Omega_{\Lambda}, \Omega_{m},\Omega_{r}\} = \{0.686, 0.314, 9.26 \times 10^{-5}\}$. 
The dots represent $z=0, 1,  5,  30, 200$, respectively, from left to right.
} 
\label{fig:f(X)}
\end{figure}

The last step to finalize the zeroth-order equations is to reconstruct the free function $f$ for a given expansion history, for which Ref. \cite{DW-2009} provides a general technique. 
 That is, one can solve a differential equation for $f$ as the following integral (see Eq. (43) of Ref. \cite{DW-2009}):
\be
f(\zeta) = -2\int_{\zeta}^{\infty} \!\! d\zetaa \,\zetaa \phi(\zetaa)
- 6 \Omega_{\Lambda} \int_{\zeta}^{\infty} \!\! d\zetaa \, \frac{\zetaa^{
2}}{h(\zetaa) I(\zetaa)} \int_{\zetaa}^{\infty} \!\! d\zetaaa \,
\frac{I(\zetaaa)}{\zetaaa^{ 4} h(\zetaaa)} 
+ 2 \int_{\zeta}^{\infty} \!\! d\zetaa \, \frac{\zetaa^{
2}}{h(\zetaa) I(\zetaa)} \int_{\zetaa}^{\infty} \!\! d\zetaaa \,
\frac{r(\zetaaa)\phi(\zetaaa)}{\zetaaa^{5}} \;. 
\eql{ffin}
\ee 
Here, the time variable $\zeta$ is defined as
$
\zeta \equiv 1 + z = 1/a,
$
and the dimensionless Hubble parameter $h$ and the dimensionless Ricci scalar $r$ are
\be
h \equiv \frac{H}{H_0}, \quad H \equiv \frac{\dot{a}}{a} 
\quad \mbox{and} \quad 
r \equiv \frac{R}{H_0^2} = 6(\dot{h} + 2h^2)  \;,
\ee
where $H_0$ is the Hubble parameter today. 
The functions $\phi(\zeta)$ and $I(\zeta)$ are \cite{DW-2009}
\be
\phi(\zeta) = -6\Omega_{\Lambda} \int_{\zeta}^{\infty} \!\! d\zetaa \,
\frac1{h(\zetaa)} \int_{\zetaa}^{\infty} \!\! d\zetaaa \, 
\frac1{\zetaaa^{4} h(\zetaaa)} \;, \quad 
I(\zeta) = \int^\infty_\zeta d\zetaa \frac{r(\zetaa)}{\zetaa^{4}h(\zetaa)}\; .
\eql{Phi_Idef}
\ee
Once an expansion history is specified in terms of $h$ as a function of $\zeta$ 
[e.g., $h^2(\zeta) = \Omega_{\Lambda} + \Omega_m \zeta^3 + \Omega_r \zeta^4$ for $\Lambda$CDM], 
one can numerically integrate \Ec{ffin} to get $f$ as a function of $\zeta$. \footnote{Note that the expansion history does not have to be that of $\Lambda$CDM. The function $f$ for a simple non-$\Lambda$CDM expansion has been evaluated in Ref. \cite{PS-2016}.} 
Next, using 
\be
\ovX(\zeta) = -\int^{\infty}_{\zeta} \frac{d\zetaa \zetaa^2}{h(\zetaa)} 
\int^{\infty}_{\zetaa} d\zetaaa \frac{r(\zetaaa)}{\zetaaa^{4} h(\zetaaa) }
=  -\int^{\infty}_{\zeta} \frac{d\zetaa \zetaa^2}{h(\zetaa)} I(\zetaa) \;,
\ee
$\zeta$ can be converted to a function of $\ovX$ and plugging it back into \Ec{ffin} gives $f$ as a function of $\ovX$.
For the case of $\Lambda$CDM, 
the numerical solution is fit well by a simple hyperbolic tangent \cite{DW-2009}. 
The authors of Ref. \cite{DW-2009} used the five-year WMAP data \cite{WMAP-5year}, 
which were the latest back then, 
$\{\Omega_{m}, \Omega_{r}\} = \{0.28, 8.5 \times 10^{-5}\}$ and $\Omega_\Lambda = 1- \Omega_m$ = 0.72, 
and obtained an analytic parametrization $f_{\rm an}$,
\be
f_{\rm an}(\ovX) = 0.245\Bigl[\tanh(0.350Y+0.032Y^2 +0.003Y^3) -1 \Bigr] \;,
\eql{deff}
\ee
where $Y \equiv \ovX + 16.5$. 
We reevaluate $f$ using the Planck 2015 values (for TT + lowP), $\{\Omega_{m}, \Omega_{r}\} = \{0.314, 9.26 \times 10^{-5}\}$ \cite{Planck2015-XIII}, which can be fitted to an analytic function
\be
f_{\rm an Planck2015}(\ovX) = 0.243\Bigl[\tanh(0.348Z+0.033Z^2 +0.005Z^3) -1 \Bigr] \;,
\eql{fan-2015Planck}
\ee
where $Z\equiv \ovX + 16.7$.
Reference \cite{NCA-2017} takes the analytic function \Ec{deff} derived in Ref. \cite{DW-2009} (which was for $\Omega_{m} = 0.28$) and uses $\Omega_{m} = 0.3$ for the rest of the computations, which can be justified because the two resulting $f(\ovX)$'s are actually very close to each other, as can be seen in Fig. \ref{fig:f(X)}. However, it should be noted that their values at $z=0$ are different by $0.02$ at $z=0$,  and the difference between the growth functions for the DW model and GR is about $0.04$ at $z=0$ (see Fig. \ref{fig:solutions-of-SetAC-GR}). Thus, for precision tests and also for consistency, one should fix $f(\ovX)$ according to the chosen expansion history.

\section{Perturbation Equations}
\label{perturbation-eq}

We take scalar perturbations around the background FLRW metric in the Newtonian gauge,
\be
g_{00}(t, \vec{x}) = -1 -2\Psi(t,\vec{x})\;, \quad
g_{0i}(t, \vec{x}) = 0 \;, \quad
g_{ij}(t, \vec{x}) = \delta_{ij} a^2(t) \Bigr[1 + 2\Phi(t,\vec{x})\Bigl]\;. 
\eql{FLRW-metric}
\ee 
Plugging the perturbed metric \Ec{FLRW-metric} into the field equations \Ec{nonlocal_field_eq} and expanding them to the first order gives the perturbation equations. 
We symbolically write the perturbation equations as
\bea
\delta\Big(G_{\mu\nu} + \Delta G_{\mu\nu} \Big) = 8\pi G \delta T_{\mu\nu}\;.
\eql{perturbation-eqns}
\eea
In our notation, little $\delta$ refers to first-order perturbations, and capital $\Delta$ to the nonlocal corrections to the original Einstein tensor. Working in Fourier space,  the Einstein tensor at first order is     
\bea
\delta G_{00} &\!\!\!=\!\!\!& 6H\dot{\Phi} + 2\frac{k^2}{a^2}\Phi \;, 
\\
\delta G_{ij} &\!\!\!=\!\!\!&  \delta_{ij}a^2\Bigl[2(-2\dot{H} - 3H^2)(\Phi - \Psi) 
+ 2H(\dot{\Psi} - 3\dot{\Phi}) - 2\ddot{\Phi} - \frac{k^2}{a^2}(\Phi + \Psi)\Bigr] +k_{i}k_{j}(\Phi + \Psi) \;.
\eea    
The modified Einstein tensor at first order is 
\bea
\delta \Delta G_{00}  &\!\!\!=\!\!\!&
\overline{\mathcal{D}}_{00}(\delta f + \delta U) + \delta \mathcal{D}_{00}(\ovf + \ovU) 
+ \frac{1}{2}\partial_t{\ovX}\partial_t{\delta U} + \frac{1}{2}\partial_t{\delta X}\partial_t{\ovU} \;,
\eql{DeltaG00}
\\
 \delta \Delta G_{ij}  &\!\!\!=\!\!\!&
 \overline{\mathcal{D}}_{ij}(\delta f + \delta U) + \delta \mathcal{D}_{ij}(\ovf + \ovU) 
+ \delta_{ij}a^2\left[(\Phi-\Psi)\partial_t{\ovX}\partial_t{\ovU} + \frac{1}{2}\partial_t{\ovX}\partial_t{\delta U} + \frac{1}{2}\partial_t{\delta X}\partial_t{\ovU}\right] \;.
\eql{DeltaGij}
\eea
Here the $00$ and $ij$ components of the operator $\mathcal{D}_{\mu\nu}$ defined in \Ec{mathcalD} are
\bea
\overline{\mathcal{D}}_{00} &\!\!\!\!=\!\!\!\!& 3H^2 + 3H\partial_{t} -\frac{\nabla^2}{a^2} \;,
\\
\delta \mathcal{D}_{00}  &\!\!\!\!=\!\!\!\!&  6H\dot{\Phi} + 3\dot{\Phi}\partial_{t} + 2\frac{k^2}{a^2}\Phi \;, 
\\
\overline{\mathcal{D}}_{ij} &\!\!\!\!=\!\!\!\!& 
\delta_{ij}a^2(-2\dot{H} - 3H^2 -2H\partial_{t}-\partial_{t}^2) + \delta_{ij}\nabla^2 -\partial_{i}\partial_{j} \;,
\\
\delta \mathcal{D}_{ij} &\!\!\!\!=\!\!\!\!& \delta_{ij}a^2\biggl\{2(-2\dot{H} - 3H^2)(\Phi - \Psi) 
+ 2H(\dot{\Psi} - 3\dot{\Phi}) -2\ddot{\Phi} + \Bigl[\dot{\Psi} - 2\dot{\Phi} - 4H(\Phi - \Psi)\Bigr]\partial_{t} 
\nn \\
& & \hspace{1.2cm}   -2(\Phi - \Psi)\partial_{t}^2 - \frac{k^2}{a^2}(\Phi + \Psi)\biggr\} + k_{i}k_{j}(\Phi + \Psi)\;.
\eea
In the derivation of $\delta \mathcal{D}_{\mu\nu}$, the terms containing spatial derivatives have been dropped because they vanish when acting on zeroth-order values (which are functions of time only):
\bea
\delta(D_0D_0) &\!\!\!=\!\!\!& -\dot{\Psi}\partial_t -\Psi\frac{ik^k}{a^2}\partial_k 
~\rightarrow~ -\dot{\Psi}\partial_t\;, 
\\
\delta(D_iD_j) &\!\!\!=\!\!\!& -\delta_{ij}a^2\Big[2H(\Phi-\Psi)+\dot{\Phi}\Big]\partial_t
+i\Phi\delta_{ij}k^k\partial_k - i\Phi(k_i\partial_j+k_j\partial_i)
~\rightarrow~ -\delta_{ij}a^2\Big[2H(\Phi-\Psi)+\dot{\Phi}\Big]\partial_t \;, 
\\
\delta \square &\!\!\!=\!\!\!& 2\Psi\partial_t^2 + (6H\Psi+\dot{\Psi}-3\dot{\Phi})\partial_t
- 2\Phi\frac{\nabla^2}{a^2} + (\Phi+\Psi)\frac{ik^k}{a^2}\partial_k 
~\rightarrow~ 2\Psi\partial_t^2 + (6H\Psi+\dot{\Psi}-3\dot{\Phi})\partial_t
\eql{deltasquare}
\eea  

We also need the expansions for $X, U,$ and $f$ at first order in $\Phi$ and $\Psi$. First, $f$ can be expanded as
\be
f(X) =  f(\ovX + \delta X) = f(\ovX) + f'(\ovX)\delta X \equiv \ovf + \delta f \;.
\ee
To get $\delta X$ recall that 
\be
0 = \delta(1) = \delta\bigg(\square \frac{1}{\square}\bigg) 
= \delta \square \frac{1}{\square} + \square\delta \frac{1}{\square}
\quad \Rightarrow \quad 
\delta \Big(\frac{1}{\square}\Big) = - \frac{1}{\square} \delta \square \frac{1}{\square} \;,
\ee
which leads to
\be
\delta X = \delta \Big(\frac{1}{\square} R \Big) = \delta \Big(\frac{1}{\square}\Big)R + \frac{1}{\square}\delta R
= - \frac{1}{\square} \delta \square \frac{1}{\square} R +  \frac{1}{\square}\delta R \;.
\ee
Here, since $\delta \square$ and $\delta R$ are already at first order, $\delta X$ at first order becomes
\be
\delta X = \frac{1}{\osq}\Big[ \delta R -  \delta \square \frac{1}{\osq} \ovr \Big] = \frac{1}{\osq}\Big[ \delta R -  \delta \square \ovX \Big]\;.
\eql{deltaX-nonlocal}
\ee
In the same way, $\delta U$ at first order is
\be
\delta U = \frac{1}{\osq}\biggl\{\delta R f'(\ovX) + \ovr f''(\ovX) \delta X  - \delta \square \frac{1}{\osq}\Big[\ovr f'(\ovX)\Big]\biggr\} = \frac{1}{\osq}\left[\delta R f'(\ovX) + \ovr f''(\ovX) \delta X  - \delta \square \ovU \right] \;.
\eql{deltaU-nonlocal}
\ee
In the localized formulation, 
the differential equations \Ec{eq:boxx} and \Ec{eq:boxu} for $X$ and $U$ can be expanded as 
\bea
\square X = R
\quad &\longrightarrow& \quad 
(\osq + \delta \square)(\ovX + \delta X) = \ovr + \delta R \;,
\nn \\
\quad &\longrightarrow& \quad 
\osq \ovX = \ovr  \quad \mbox{and} \quad 
\osq \delta X + \delta \square \ovX = \delta R \;, 
\eql{expandX-local}
\\
\square U = R f'(X)
\quad &\longrightarrow& \quad 
(\osq + \delta \square)(\ovU + \delta U) = (\ovr + \delta R)\Big[f'(\ovX) + f''(\ovX)\delta X \Big] \;,
\nn \\
\quad &\longrightarrow& \quad 
\osq \ovU = \ovr f'(\ovX)  \quad \mbox{and} \quad 
\osq \delta U + \delta \square \ovU =  \delta R f'(\ovX) + \ovr  f''(\ovX)\delta X \;.
\eql{expandU-local}
\eea
These local auxiliary equations at first order are equivalent to the corresponding nonlocal expressions 
\Ec{deltaX-nonlocal} and \Ec{deltaU-nonlocal},
as they should be,
\bea
\osq \delta X  &=& \delta R - \delta \square \ovX\;, 
\eql{deltaX-local}
\\
\osq \delta U  &=&  \delta R f'(\ovX) + \ovr  f''(\ovX)\delta X - \delta \square \ovU\;.
\eql{deltaU-local}
\eea 

This fulfills 
the linear perturbation equations \Ec{perturbation-eqns} for the gravitational potentials $\Phi$ and $\Psi$ 
with all the components $\delta G_{00}, \delta \Delta G_{00}, \delta G_{ij}$,  and $\delta \Delta G_{ij}$ and the two auxiliary equations for $\delta X$ and $\delta U$ given above.  
For the $ij$ equations, contracting with the projector operator $\hat{k}^i\hat{k}^j-1/3\delta^{ij}$ greatly simplifies them by extracting the longitudinal traceless component,
\be
\Big(\hat{k}^i\hat{k}^j-\frac{1}{3}\delta^{ij}\Big)\Big[\delta G_{ij} + \delta \Delta G_{ij}\Big]
= \Big(\hat{k}^i\hat{k}^j-\frac{1}{3}\delta^{ij}\Big) 8\pi G \delta T_{ij} = -8\pi G  (\rho+p)  
\sigma\;,
\ee 
where $\sigma$ represents the anisotropic stress in the convention of Ref. \cite{Ma-Bert1995}. The linear perturbation equations for $\Phi$ and $\Psi$ are then explicitly written as 
\bea
\lefteqn{
6H\dot{\Phi} + 2\frac{k^2}{a^2}\Phi + 
\Big(3H^2 + 3H\partial_{t} +\frac{k^2}{a^2} \Big)( \ovf' \delta X+ \delta U) + 
\Big(6H\dot{\Phi} + 3\dot{\Phi}\partial_{t} + 2\frac{k^2}{a^2}\Phi \Big)(\ovf + \ovU) 
+ \frac{1}{2}\Big(\partial_t{\ovX}\partial_t{\delta U} + \partial_t{\delta X}\partial_t{\ovU}\Big)}
\nn \\
&& = 8\pi G\rho 
\delta\;, 
\eql{full-linear-00-eq}
\qquad \qquad \qquad \qquad \qquad \qquad \qquad \qquad \qquad \qquad \qquad \qquad
\qquad \qquad \qquad \qquad \qquad \qquad \qquad \quad \quad \quad \quad \quad
\eea
where $\delta T_{00} = \rho\delta$ and $\delta = \delta \rho/\rho$ is the matter overdensity, and 
\be
\frac{2}{3}k^2(\Phi+\Psi) +\frac{2}{3}k^2 (\ovf' \delta X  + \delta U) + \frac{2}{3}k^2(\Phi+\Psi)(\ovf + \ovU) =  -8\pi G(\rho+p) a^2 \sigma\;. 
\eql{full-linear-traceless-eq}
\ee
So far, we have not taken any approximations yet, and thus these two equations \Ec{full-linear-00-eq} and \Ec{full-linear-traceless-eq} are full linear perturbation equations, which we have not provided in our previous papers \cite{PD-2012, DP-2013}. \footnote{Equation (3.8) of Ref. \cite{NCA-2017} corresponds to \Ec{full-linear-00-eq}; however, the authors omitted some terms in the sub-horizon limit. This limit will be discussed in the next subsection. Equation (3.17) of Ref. \cite{NCA-2017} is the same as \Ec{full-linear-traceless-eq}.} 

At late times, which is the regime in which we are interested, the contributions to the energy from relativistic particles are negligible, and hence we can set $\sigma = 0$:   
\be
(\Phi+\Psi) +(\ovf' \delta X  + \delta U) + (\Phi+\Psi)(\ovf + \ovU) =  0\;. 
\eql{modified-gslip-eq}
\ee

We now have four equations for five unknown variables $\Phi,  \Psi,  \delta X,  \delta U$, and $\delta$; thus we need one more equation 
to complete the set of equations. It can be supplied by the energy-momentum conservation ($\nabla^{\mu}T_{\mu\nu} = 0$), which is guaranteed by the conservation of the modified Einstein tensor ($\nabla^{\mu}\Delta G_{\mu\nu} = 0$). 

In the next subsection, we will take the sub-horizon limit ($k \gg Ha$) since these are the scales most relevant to structure formation and then solve those five equations. However, we must emphasize here that those 
equations can be solved numerically without taking any approximation. Taking the sub-horizon limit makes the equations simpler, and 
the solutions of the full linear equations and those of the sub-horizon limit equations have a negligible difference (for instance, in the scales of $k=100H_0$, the difference is order of $10^{-4}$),  
which justifies taking the limit. 
However, there are important subtleties to be considered when taking this limit, which we will discuss next in detail.

\subsection{Sub-horizon limit}

An additional advantage of taking the sub-horizon limit is that Green's function for the differential operator $\osq$ in the two auxiliary equations \Ec{deltaX-local} and \Ec{deltaU-local} has an analytic solution in that limit.
The retarded Green's function for the operator $\osq$ satisfies 
\be
\overline{\square} G_{ret}(x;x') = 
\Bigl(-\partial_t^2 - 3H \partial_t +\frac{\nabla^2}{a^2} \Bigr)G_{ret}(x;x') = \delta^4(x-x')\;.
\ee 
This Green's function can be constructed using the massless, minimally coupled scalar mode functions $u(t,k)$: 
\be
G_{ret}(x;x') = \frac{\Theta(t-t')a^3(t')}{i} 
\int \frac{d^3k}{(2\pi)^3}e^{i\vec{k}\cdot (\vec{x} - \vec{x'})}\Bigr[ u(t,k)u^*(t',k) - u^*(t,k)u(t',k) \Bigl]\;.
\eql{Green-4d}
\ee
There is no general solution $u(t,k)$ for an arbitrary $a(t)$, so one numerically solves for $u(t,k)$. However,  for the case of sub-horizon modes ($k \gg Ha $), we can find an analytic form for $u(t,k)$ using the Wentzel-Kramers-Brillouin (WKB) approximation,
\be
u(t,k) = \frac{1}{\sqrt{2k}}\frac{\exp\left[-ik\int^{t}\frac{dt'}{a(t')}\right]}{a(t)} \;.
\ee
Plugging these mode functions into \Ec{Green-4d} leads to the Fourier space Green's function, 
\be
G(t,t';k) = -\frac{\Theta(t-t')a^2(t')}{ka(t)}\sin\Bigl[k\int_{t'}^{t}\frac{dt''}{a(t'')}\Bigr]\;.
\ee
The two auxiliary equations, \Ec{deltaX-local} and \Ec{deltaU-local} can then be solved as 
\bea
\delta X(\vec k,t) = \int_{t_i}^{t} dt' G(t,t';k)\left[ \delta R - \delta \square \ovX \right](\vec k,t') \;,
\eql{deltaX-sol} \\
\delta U(\vec k,t) = \int_{t_i}^{t} dt' G(t,t';k)\left[ \delta R f'(\ovX) + \ovr  f''(\ovX)\delta X - \delta \square \ovU \right](\vec k,t') \;.
\eql{deltaU-sol}
\eea

Now, we take the sub-horizon limit, which practically means dropping time derivative terms (hence not having $k^2$) because they are a factor of $(H/k)^2$ smaller than the terms having $k^2$ in the scales of $k \gg Ha$. 
In this limit, the $00$ equation \Ec{full-linear-00-eq} becomes the so-called modified Poisson equation,
\be
k^2\Phi +  k^2\left[\Phi(\ovf + \ovU) + \frac{1}{2}(\ovf' \delta X + \delta U)\right]  
= 4\pi G a^2 \rho 
\delta \;.
\eql{modified-Poisson-eq}
\ee

The energy-momentum conservation $\nabla^{\mu}T_{\mu\nu} = 0$ leads to
\be
\ddot{\delta} + 2H\dot{\delta} = -\frac{k^2}{a^2}\Psi \;,
\eql{delta-evolution-eq}
\ee
which is also derived in most cosmology textbooks (see, for example, Refs. \cite{Dodelson-book} and \cite{Amendola-Tsujikawa-book}). 
Note that in this equation for $\delta$ the sub-horizon limit
is taken only on the rhs.
We do not drop the time derivatives of $\delta$ on the lhs because we are eventually interested in the time evolution of $\delta$. 
This also indicates
the different nature of the gravitational potentials $\Phi$ and $\Psi$ from that of the matter density field $\delta$. The gravitational potentials, which are parts of metric perturbations governing the geometry of the Universe, do not change much in the sub-horizon scale (while still in the linear regime).  
This is indeed the reason why we can safely 
drop the time derivatives of $\Phi$ and $\Psi$ and why the sub-horizon limit is also called the quasi-static limit. However, as one can see from \Ec{delta-evolution-eq}, 
the gravitational potentials source the matter density to grow.

How about the auxiliary fields $\delta X$ and $\delta U$? 
The authors of Ref. \cite{NCA-2017} dropped their time derivatives, assuming they are slowly varying in the sub-horizon scale. However, the authors of Ref. \cite{DP-2013} did not drop all of their time derivatives because these fields were newly introduced variables in the DW model and their dynamics were unknown.
In fact, the structure of the auxiliary equations is very similar to that of the equation for $\delta$:
\bea
\Bigl(-\partial_t^2 - 3H \partial_t -\frac{k^2}{a^2} \Bigr) \delta X &\!\!\!=\!\!\!& S_{\delta X}(\vec{k}, t)\;, \\
\Bigl(-\partial_t^2 - 3H \partial_t -\frac{k^2}{a^2} \Bigr) \delta U &\!\!\!=\!\!\!& S_{\delta U}(\vec{k}, t)\;, \\
(\partial_t^2 + 2H \partial_t)\delta &\!\!\!=\!\!\!& S_{\delta} (\vec{k}, t)\;.
\eea
In all three equations, the gravitational potentials $\Phi$ and $\Psi$ source the time evolution of $\delta X,  \delta U$, and $\delta$. 
As can be seen from \ec{delta-evolution-eq}, one takes the quasi-static limit (i.e., dropping times derivatives) for the source term on the right-hand side when interested in the dynamics of the variable on the left-hand side. The authors of Ref. \cite{DP-2013} followed this approach because they were interested in the dynamics of the nonlocal pieces $\delta X$ and $\delta U$. The objective of their study after all was to find out how the nonlocal modification (which is encoded in $\delta X$ and $\delta U$) affects geometry and eventually matter. 

Taking the sub-horizon limit in the source term of the auxiliary field equations leads to
\bea
S_{\delta X}
&\!\!\!=\!\!\!& \delta R - \delta \square \ovX 
~ \rightarrow ~ \delta R\;, 
\\
S_{\delta U}
&\!\!\!=\!\!\!&  \delta R f'(\ovX) + \ovr  f''(\ovX)\delta X - \delta \square \ovU 
= \delta R f'(\ovX) + \ovr  f''(\ovX)\frac{1}{\osq}\left(\delta R - \delta \square \ovX\right) - \delta \square \ovU 
~ \rightarrow ~  \delta R f'(\ovX)\;.
\eql{deltaU-local-sub-horizon}
\eea 
Here, $\delta \square ~\rightarrow~ 0$ from \ec{deltasquare} and the second term of $S_{\delta U}$ (which is $\osq^{-1}\delta R \sim k^{-2}k^2 = k^0$) drop in the sub-horizon limit. 
Then, the auxiliary equations simply become
\bea
\osq \delta X &\!\!\!=\!\!\!& \delta R 
\quad \mbox{or, explicitly,} \quad  
\Bigl(-\partial_t^2 - 3H \partial_t -\frac{k^2}{a^2} \Bigr) \delta X =  2\frac{k^2}{a^2}(\Psi+2\Phi)  \;,
\eql{deltaX-local-sub-horizon}
\\
\osq \delta U &\!\!\!=\!\!\!&  \delta R f'(\ovX) 
\quad \mbox{or, explicitly,} \quad  
\Bigl(-\partial_t^2 - 3H \partial_t -\frac{k^2}{a^2} \Bigr) \delta U =  2\frac{k^2}{a^2}(\Psi+2\Phi)f'(\ovX)  \;.
\eql{deltaU-local-sub-horizon}
\eea 

Instead, if the quasi-static limit had also been taken on the left-hand side, they would have become 
\bea
\delta X &\!=\!& -2(\Psi+2\Phi)\;,  
\label{deltaX-NCA}
\\
\delta U &\!=\!& -2f'(\ovX)(\Psi+2\Phi)\;,
\label{deltaU-NCA}
\eea
which are exactly the same as Eqs. (3.19) and (3.20) in Ref. \cite{NCA-2017}. 
It should be noted that $\delta X$ and $\delta U$ are now local in this implementation of the quasi-static limit. 
We recall that $X$ was originally nonlocal, as one can see from $X = \square^{-1}R$, and this is why the DW model is called nonlocal. This is another reason why the authors of Ref. \cite{DP-2013} were reluctant to drop the time derivatives of $\delta X$.

In summary, we have a system of five equations for five perturbation variables, $\Phi, \Psi, \delta, \delta X$, and $\delta U$ in the localized formulation:
\bea
k^2\Phi +  k^2\left[\Phi(\ovf + \ovU) + \frac{1}{2}( 
\ovf' \delta X + \delta U)\right]  &\!=\!& 4\pi G a^2 \rho 
\delta \;,
\label{modified-Poisson-eq-summary}
\\
(\Phi+\Psi)+ (\ovf' \delta X + \delta U)+(\Phi+\Psi)(\ovf + \ovU) &\!=\!& 0\;,
\label{modified-gslip-eq-summary}
\\
\ddot{\delta} + 2H\dot{\delta} &\!=\!& -\frac{k^2}{a^2}\Psi \;,
\label{delta-evolution-eq-summary}
\\
\Bigl(-\partial_t^2 - 3H \partial_t -\frac{k^2}{a^2} \Bigr) \delta X &\!=\!&  2\frac{k^2}{a^2}(\Psi+2\Phi)\;,
\label{deltaX-local-sub-horizon-summary}
\\
\Bigl(-\partial_t^2 - 3H \partial_t -\frac{k^2}{a^2} \Bigr) \delta U &\!=\!&  2\frac{k^2}{a^2}(\Psi+2\Phi)
\ovf' \;.
\label{deltaU-local-sub-horizon-summary}
\eea
Or, equivalently, inserting the integral solutions for $\delta X$ and $\delta U$, 
\bea
\delta X(\vec k,t) &\!=\!& \int_{t_i}^{t} dt' G(t,t';k) \frac{2k^2}{a(t')^2}\left[\Psi(\vec{k},t')+2\Phi(\vec{k},t') \right] \;,
\\
\delta U(\vec k,t) &\!=\!& \int_{t_i}^{t} dt' G(t,t';k)\frac{2k^2}{a(t')^2}\left[\Psi(\vec{k},t')+2\Phi(\vec{k},t') \right]f'(\ovX(t')) \;,
\eea
leads to a system of three equations for three variables $\Phi, \Psi$, and $\delta$:
\bea
k^2\Phi +  k^2\left\{ \Phi(\ovf + \ovU) 
+k^2\int_{t_i}^{t}\!\!\frac{dt'}{a^2(t')}G(t,t';k)\left[\Psi(\vec{k},t')+2\Phi(\vec{k},t') \right]\!\!
\left[ f'(\ovX(t))+f'(\ovX(t'))\right] \!
\right\}
&\!\!=\!\!& 4\pi G a^2 \rho \delta \;,
\label{modified-Poisson-eq-original}
\\
(\Phi+\Psi) +(\Phi+\Psi)(\ovf + \ovU)
+2k^2\int_{t_i}^{t}\!\!\frac{dt'}{a^2(t')}G(t,t';k)\left[\Psi(\vec{k},t')+2\Phi(\vec{k},t') \right]\!\!
\left[ f'(\ovX(t))+f'(\ovX(t'))\right] 
&\!\!=\!\!& 
0\;,
\label{modified-gslip-eq-original}
\\
\ddot{\delta} + 2H\dot{\delta} 
&\!\!=\!\!& 
 -\frac{k^2}{a^2}\Psi \;,
\label{delta-evolution-eq-original}
\eea
which was used by the authors of Ref. \cite{DP-2013}.
As long as the initial conditions are taken the same, these two sets of equations should give the same answers for $\Phi, \,\Psi,$ and $\delta$, respectively.

\subsection{Solutions of the three sets of perturbation equations}

\begin{figure*}[!t] 
\begin{center}
 \begin{tabular}{cc} 
  \includegraphics[width=0.45\textwidth]{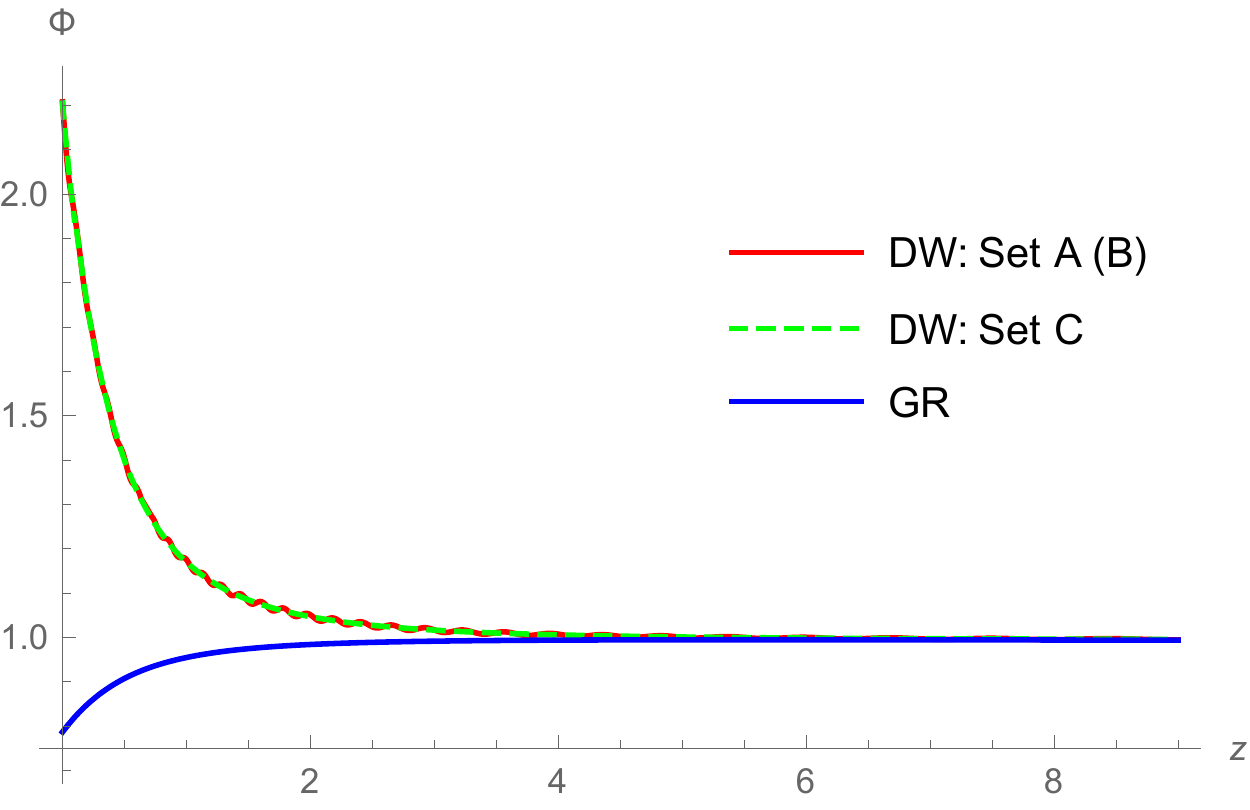}
  &
  \includegraphics[width=0.45\textwidth]{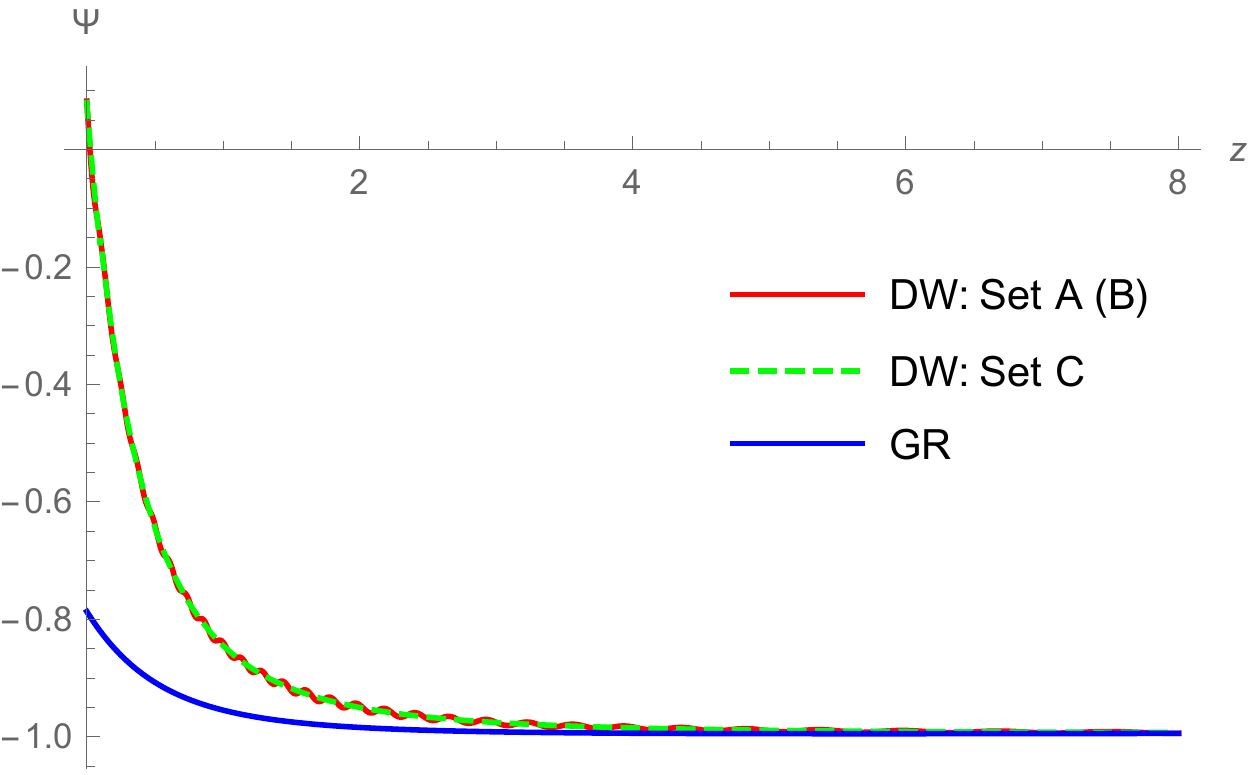}
\end{tabular}
\end{center}
\begin{center}
 \begin{tabular}{cc} 
  \includegraphics[width=0.48\textwidth]{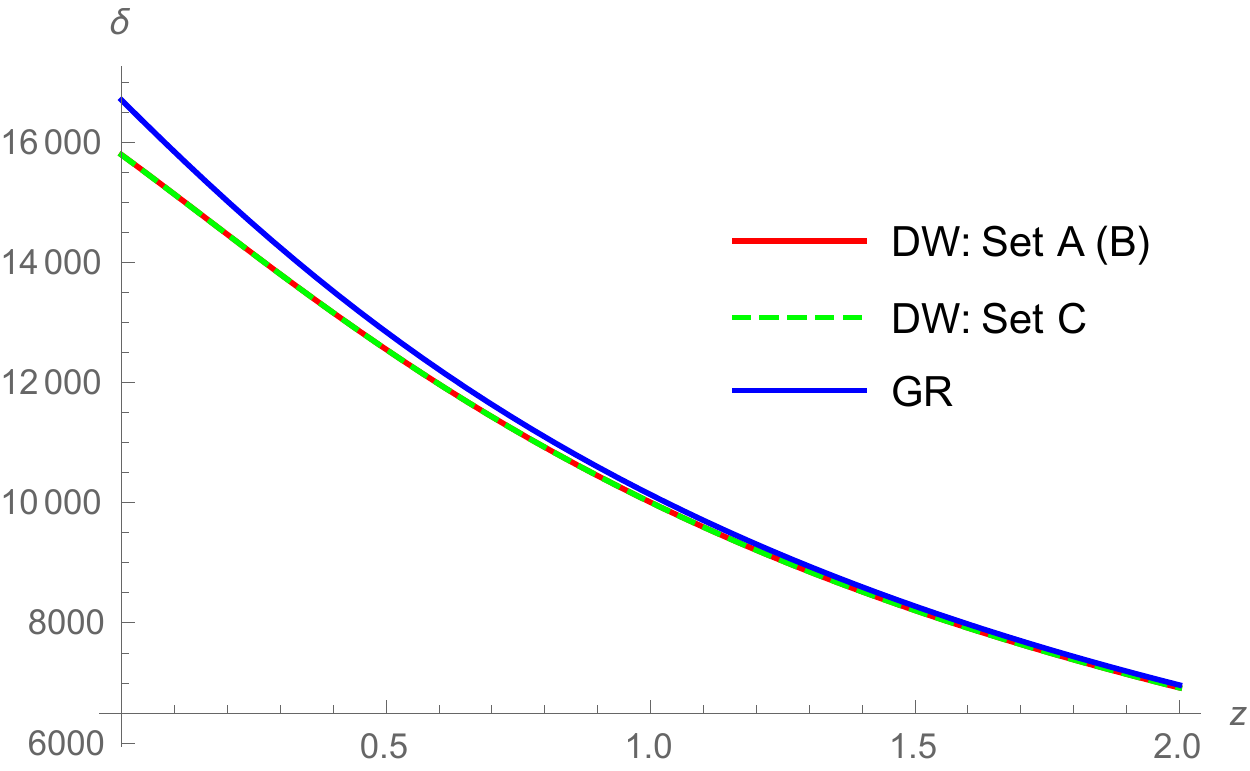} 
  &
  \includegraphics[width=0.47\textwidth]{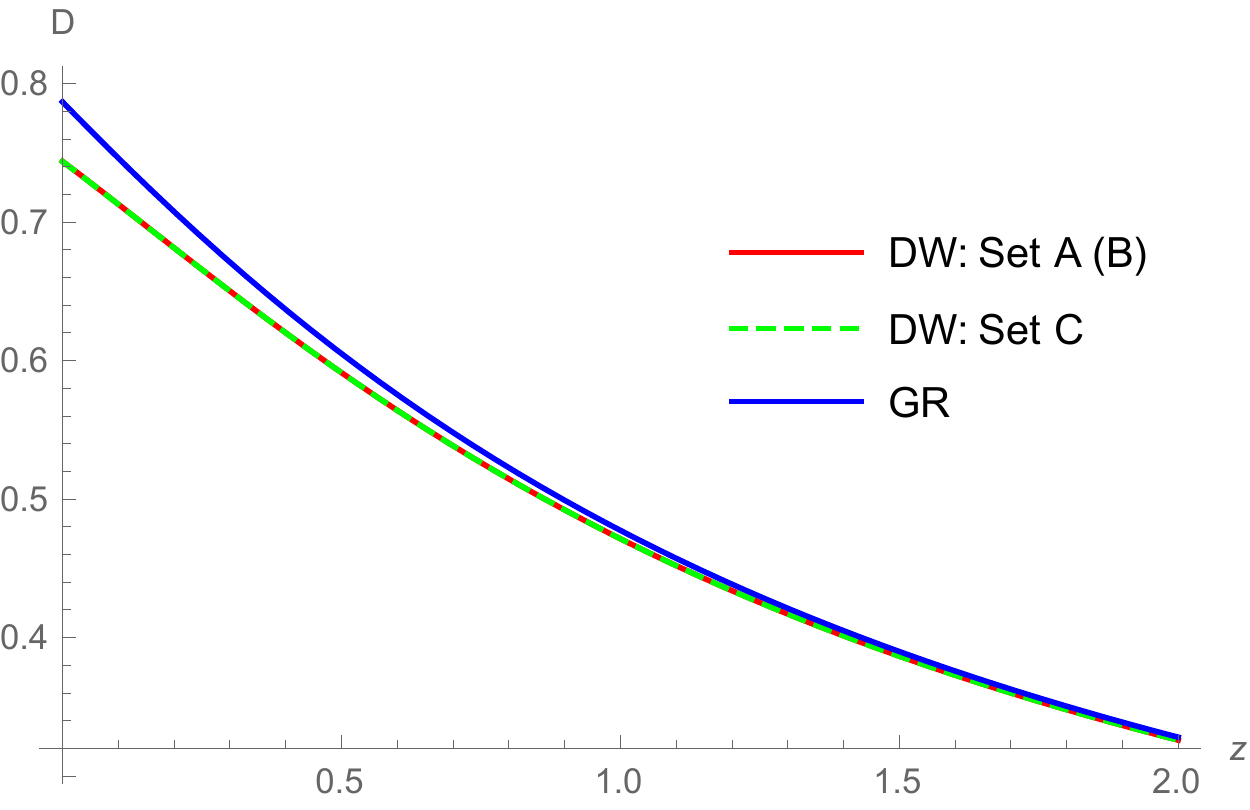} 
 \end{tabular}
\end{center}
\begin{center}
 \begin{tabular}{cc} 
  \includegraphics[width=0.45\textwidth]{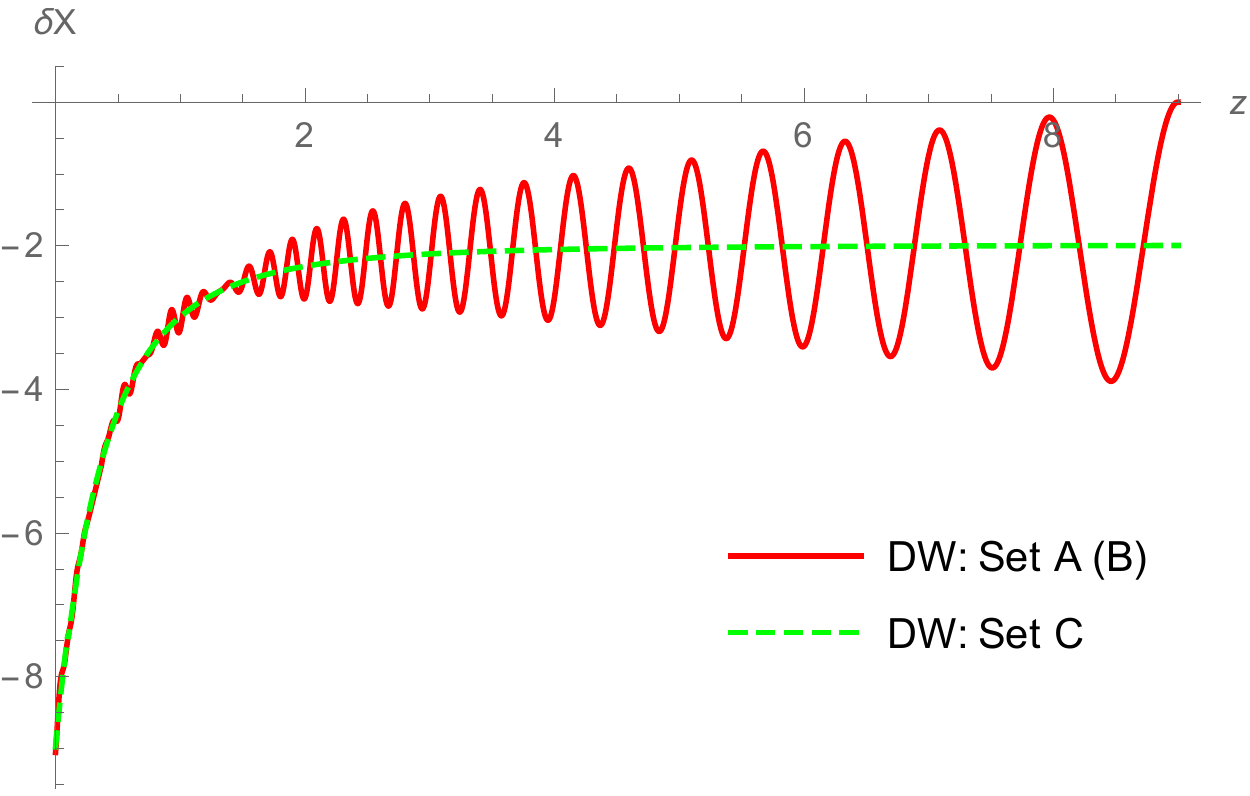}
  &
  \includegraphics[width=0.45\textwidth]{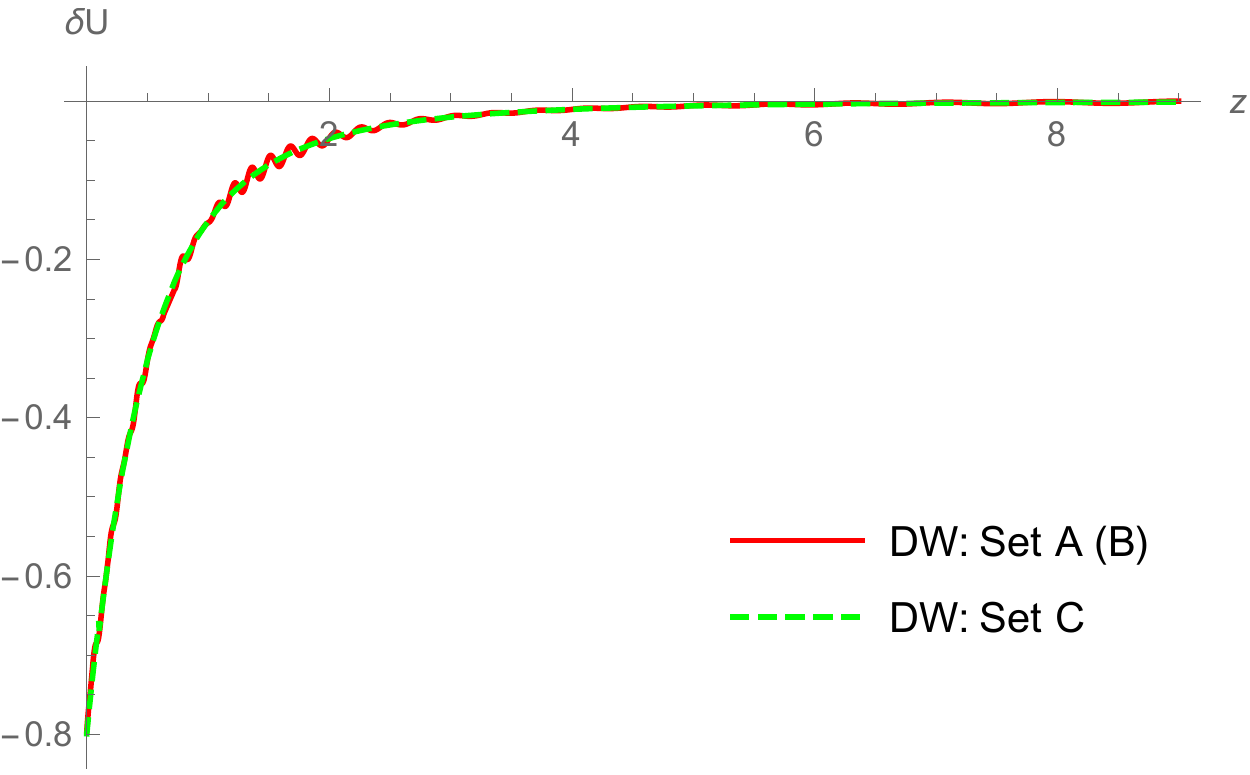}
 \end{tabular}
\end{center}
\caption{The solutions of set A (or set B), set C, and GR for $\Phi(z), \Psi(z), \delta(z)$, and $D(z)$ and of sets A and C for $\delta X(z)$ and $\delta U(z)$.} 
\label{fig:solutions-of-SetAC-GR}
\end{figure*}

To compare with the results in Ref. \cite{NCA-2017}, we solve the three sets of perturbation equations:
\begin{itemize}
\item{Set A: five equations (\ref{modified-Poisson-eq-summary} - \ref{deltaU-local-sub-horizon-summary});}
\item{Set B: three equations (\ref{modified-Poisson-eq-original} - \ref{delta-evolution-eq-original}) taken by Ref. \cite{DP-2013};}
\item{Set C: five equations (\ref{modified-Poisson-eq-summary} - \ref{delta-evolution-eq-summary}) and (\ref{deltaX-NCA}, \ref{deltaU-NCA}) taken by Ref. \cite{NCA-2017}}
\end{itemize}
Set A and Set B give identical solutions for $\Phi, \Psi, \delta, \delta X$, and $\delta U$, as they should.  Setting initial conditions at $z_i = 9$ the same as in GR (because the nonlocal modification is negligible at $z > 5$, as can be seen from Fig. 1),
\bea
\Phi(z_i) &\!\!\!=\!\!\!& \Phi_{\rm GR}(z_i)\;, ~
\Psi(z_i) = \Psi_{\rm GR}(z_i) = -\Phi_{\rm GR}(z_i)\;, \\ 
\delta(z_i) &\!\!\!=\!\!\!& \delta_{\rm GR}(z_i) = \frac{2k^2 a(z_i)}{3H_0^2 \Omega_{m}}\Phi_{\rm GR}(z_i)\;, ~ 
\delta'(z_i) = \delta'_{\rm GR}(z_i)\;,\\ 
\delta X(z_i) &\!\!\!=\!\!\!& 0\;, ~ \delta U(z_i) = 0\;, ~\delta X'(z_i) = 0\;, ~ \delta U'(z_i) = 0 \;.  
\eea
with $\Omega_m = 0.314, k=100H_0 = 0.03h \rm{Mpc}^{-1}$, we have 
numerical solutions\footnote{
Set A can be solved using  a {\it Mathematica} function {\it NDSolve}. 
For set B we discretized the equations and iterated them $10^4$ times. While the running time of set A is a few minutes, it takes set B about 11 hours with a relatively new computer. Back in 2013, running set B took days, so the current author made a simplification (a sort of localization) in which she made a coding mistake, which led to the result that
now turns out to be wrong, namely, the enhanced growth of $\delta$.} 
plotted in red in Fig. \ref{fig:solutions-of-SetAC-GR}. Their numerical values have so little difference that the two pairs of graphs from set A and set B look identical. Thus, we only provide the plots for the solutions of set A.

It should be noted that
$\delta X$ and $\delta U$ are forced (and lightly damped) harmonic oscillators with the driving forces $-2(\Psi+2\Phi)$ and $-2f'(\ovX)(\Psi+2\Phi)$ for $\delta X$ and $\delta U$, respectively, so that they will eventually follow the driving forces after transient oscillations. 
These driving forces are actually the solutions of $\delta X$ and $\delta U$ in (\ref{deltaX-NCA}) and (\ref{deltaU-NCA}) taken by Ref. \cite{NCA-2017}, and this is why the solutions of set C are asymptotically the same as the ones of set A (and set B). This behavior is verified in Fig. \ref{fig:solutions-of-SetAC-GR}, which plots the solutions of set A, set C, and GR for comparison. Because $\delta X$ is multiplied by $\ovf'$ 
(which is almost zero at high redshifts $z \gtrsim 5$) 
in Eqs. (\ref{modified-Poisson-eq-summary}) and (\ref{modified-gslip-eq-summary}), its oscillation is killed at high redshifts.
Hence, the solutions for $\Phi(z), \Psi(z)$, and $\delta(z)$ of sets A and C are almost the same. 
As pointed out in Ref. \cite{NCA-2017}, the growth of $\delta$ in the DW model is lower than in GR. 
In Fig. \ref{fig:solutions-of-SetAC-GR}, we also provide the normalized solution for $\delta$, the {\it growth function} $D(z)$,
which can be obtained by
\be
D(z) = \frac{3\Omega_m H_0^2}{2k^2} \delta(z),
\ee
or, equivalently, by solving the growth equation \Ec{growth-eq-mu} given in the next subsection  
with the same initial conditions as in Ref. \cite{NCA-2017}:
\be
D(z_i) = a(z_i)\;, \quad D'(z_i) = -D(z_i) a(z_i)\;.
\ee

\begin{figure*}[htbp] 
\begin{center}
 \begin{tabular}{cc} 
  \includegraphics[width=0.47\textwidth]{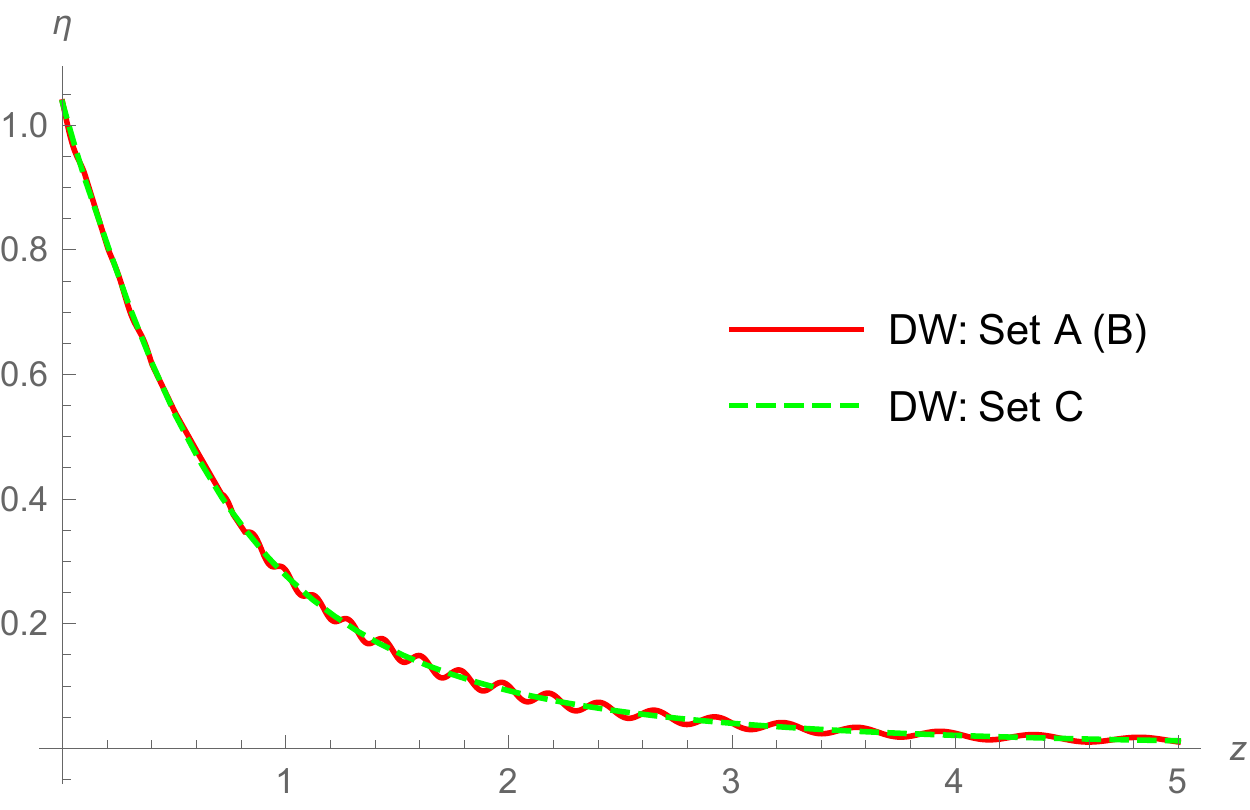}
  &
  \includegraphics[width=0.47\textwidth]{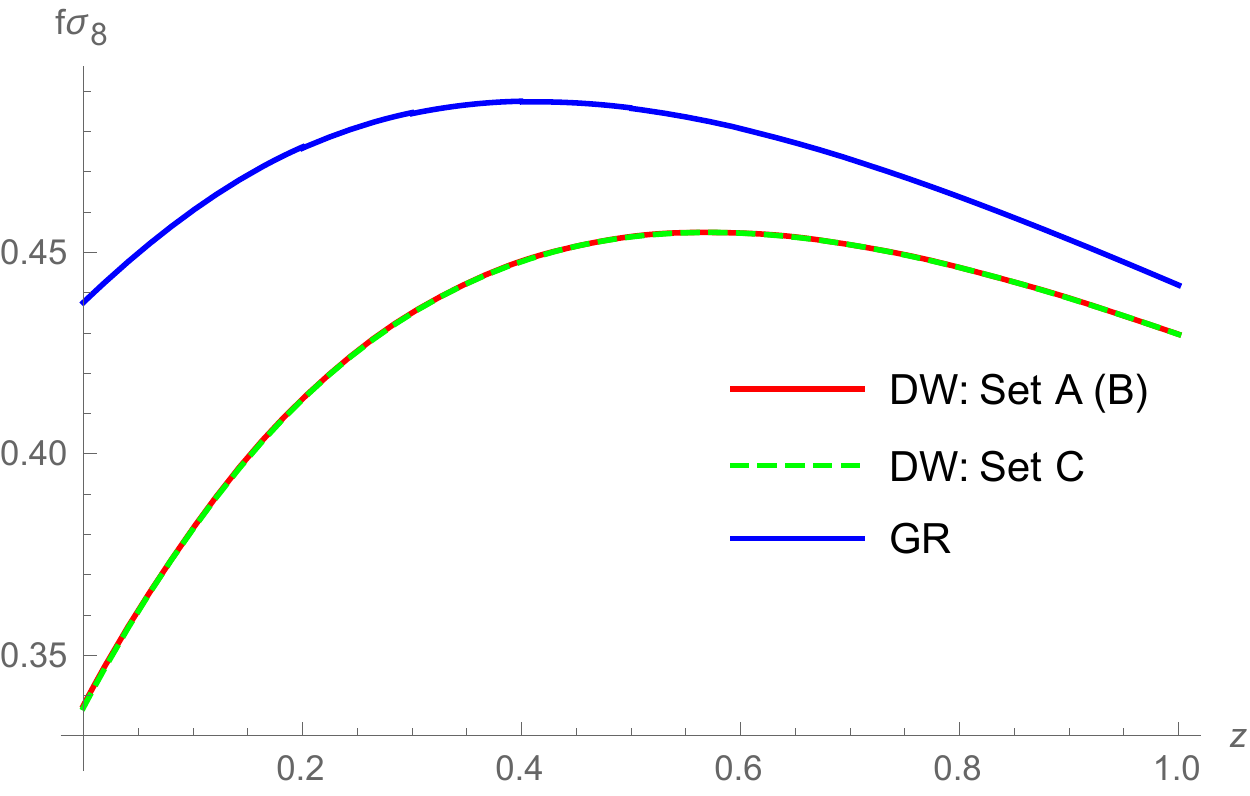}
 \end{tabular}
\end{center}
\caption{Left: The gravitational slip $\eta$ computed using the solutions for $\Phi$ and $\Psi$ from set A (or set B) and set C as a function of redshift. Right: The growth rate $f\sigma_8 \equiv d \ln \delta/d \ln(a) \times \sigma_8(z) $ using the solutions for $\delta$ from set A (or set B), set C, and GR as a function of redshift.} 
\label{fig:eta-fsigma8}
\end{figure*}

Figure \ref{fig:eta-fsigma8} depicts the gravitational slip $\eta  \equiv (\Phi + \Psi)/\Phi $ and the growth rate $f\sigma_8 \equiv d \ln \delta/d \ln(a) \times \sigma_8(z) $, which are in fair agreement with the ones obtained in Ref. \cite{NCA-2017}.
The slight difference  in the numerical values of $\eta$ in Fig. \ref{fig:eta-fsigma8} and in Fig. 1 of Ref. \cite{NCA-2017} might be due to their different choices of the nonlocal distortion function $f$ (which cause their background solutions for $\ovX$ and $\ovU$, and hence $\Phi$ and $\Psi$, to differ slightly from the ones in this manuscript) and different $\Omega_m$ values.
The normalization condition for $\sigma_8(z)$ is taken to be the one given in Ref. \cite{DP-2013}, which is setting the initial amplitude $\sigma_8(z_i)$ the same for GR, i.e., $\sigma_8(z_i) = \sigma_8(z=0) \frac{\delta_{\rm GR}(z_i)}{\delta_{\rm GR}(z=0)}$. 
Reference \cite{NCA-2017} used a slightly different normalization condition that also sets  the initial amplitude  identical to GR but evolves it with the solution for $\delta$ of the DW model.
Also, while we set $ \sigma_8(z=0) = 0.83$ according to the Planck 2015 results \cite{Planck2015-XIII},  Ref. \cite{NCA-2017} sets $ \sigma_8(z=0) = 0.78$.

\subsection{Alternative definitions of the effective gravitational constant $G_{\rm eff}$}

One confusion between Refs. \cite{DP-2013} and \cite{NCA-2017} regarding 
the modified strength of gravity might arise from the alternative definitions of 
the effective gravitational constant $G_{\rm eff}$ (in units of $G_{\rm Newton}$).  In Ref. \cite{DP-2013}, it was defined through the relation between $\Phi$ and $\delta$ (the modified Poisson equation), 
whereas in Ref. \cite{NCA-2017}, it was defined through the relation between $\Psi$ and $\delta$:
\be
\frac{k^2}{a^2}\Phi \equiv G_{\rm{eff}  \Phi } \times 4\pi G 
\rho \delta  \quad \mbox{vs.} \quad 
-\frac{k^2}{a^2}\Psi \equiv G_{\rm{eff} \Psi } \times 4\pi G 
\rho \delta \;.
\ee
If the gravitational slip $\eta = (\Phi + \Psi)/\Phi $ vanished (at late times) like in GR, these two definitions would be identical; however, it is not the case of the DW nonlocal model. 
In fact, these two $G_{\rm eff}$'s behave in opposite ways. While $G_{\rm{eff} \Phi}$ increases with time, $G_{\rm{eff} \Psi}$ decreases. That is, while $|\Phi|$ in the DW model becomes larger than in GR, $|\Psi|$ becomes smaller. This is in contrast to the case of GR in which $|\Phi| = |\Psi|$. Another way to see the difference is that while $G_{\rm{eff} \Phi}$ corresponds to $G_{\rm{eff}}$ in the background level,  $G_{\rm{eff} \Psi}$ more directly characterizes the growth of $\delta$. 
 
\begin{figure*}[htbp] 
\begin{center}
 \begin{tabular}{cc} 
  \includegraphics[width=0.47\textwidth]{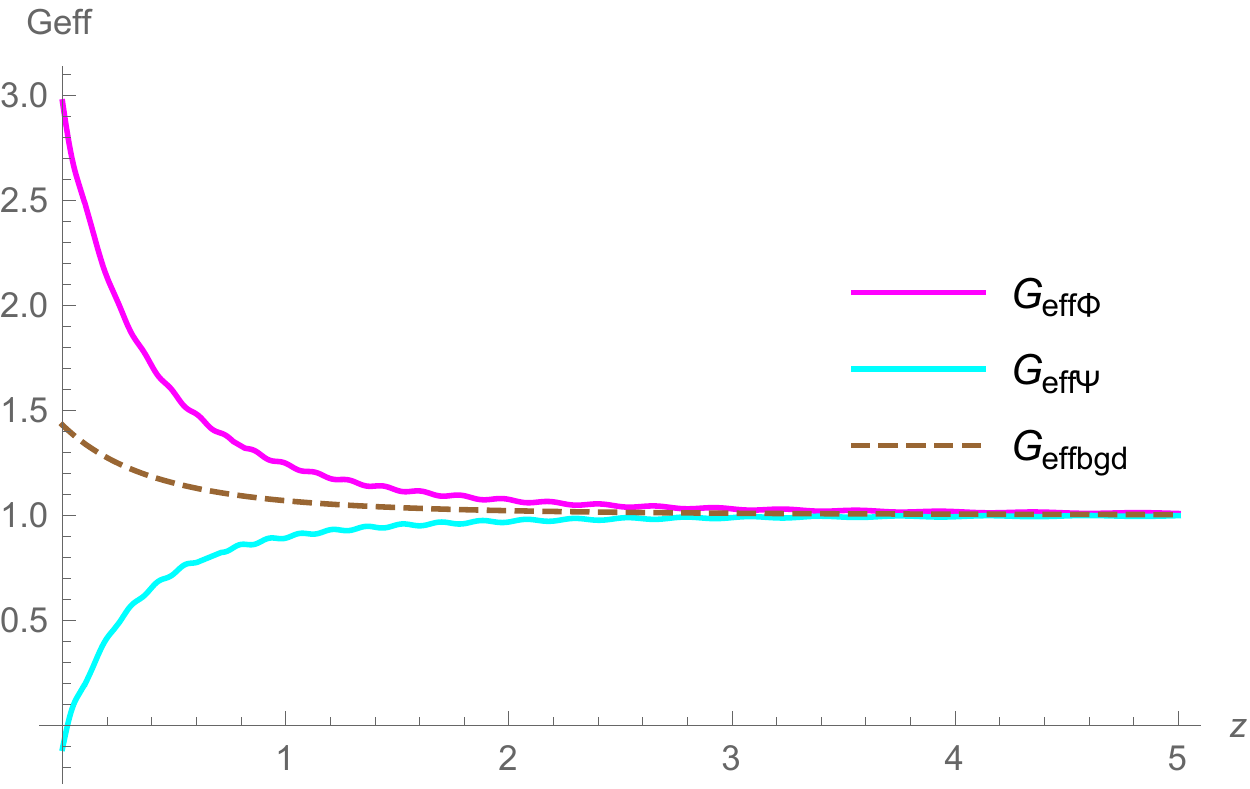}
  &
  \includegraphics[width=0.47\textwidth]{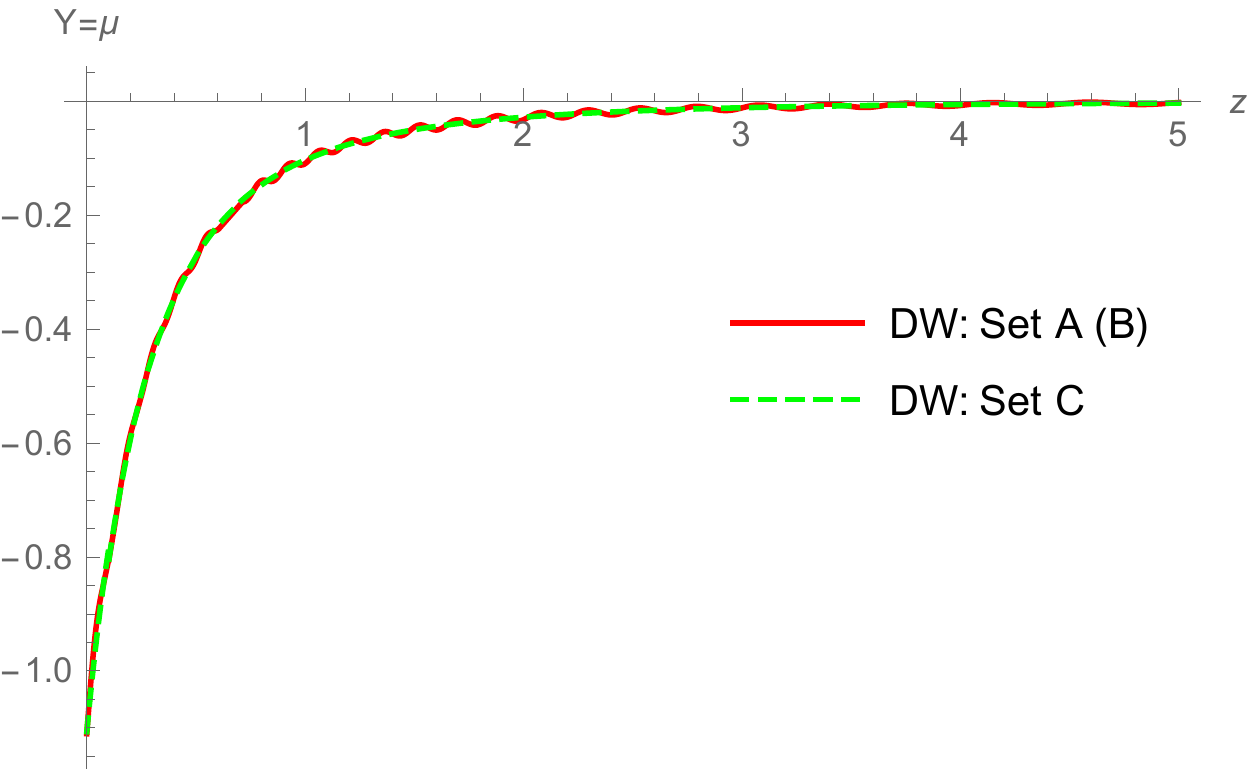}
 \end{tabular}
\end{center}
\caption{Left: Alternative ways of defining the effective gravitational constant $G_{\rm{eff}}$:  $G_{\rm{eff} \Phi}$ from the relation between $\Phi$ and $\delta$,  $G_{\rm{eff} \Psi}$ from the relation between $\Psi$ and $\delta$, and $G_{\rm{eff bgd}}$ from the modified Friedmann equation. 
The curves are wiggly because we used the solutions of set A. Instead, if one uses the solutions of set C, the curves are smooth with the same overall shapes.
Right: The parameter Y defined in \Ec{growth-eq-Y}, which turns out to be the same as the parameter $\mu$ defined in \Ec{mu-def}: Their negative sign indicates the lower growth of $\delta$ compared to the one in GR.} 
\label{fig:Geff-Y=mu}
\end{figure*}

In the background level, what the DW model essentially does is increase  $G_{\rm{eff}}$ in order to overcome the falling-down matter density $\rho$. This can be seen from the Friedmann equation without $\Lambda$: 
\be
H^2 = \frac{8\pi G}{3} \rho\;.
\ee 
The problem is that, while the left-hand side is observed to be approaching
a constant, the matter density in the right-hand side is dropping as $a^{-3}$. 
The DW model modifies the Friedmann equation, thereby increasing the gravitational constant so that the left- and right-hand sides match each other:
\be
H^2 = \frac{8\pi G \times G_{\rm{eff}}}{3} \rho\;.
\ee 
The left panel of Fig. \ref{fig:Geff-Y=mu} 
depicts  $G_{\rm{eff} \Phi}, G_{\rm{eff} \Psi}$, and $G_{\rm{eff bgd}}$ ($G_{\rm{eff}}$ in the background level), where $G_{\rm{eff bgd}}$ is obtained from $G_{\rm{eff} \Phi}$ by taking the background modifications only,
\be
G_{\rm{eff} \Phi} = \frac{1}{1 + \ovf + \ovU +\frac{1}{2\Phi} (\ovf' \delta X + \delta U)} 
\quad \longrightarrow \quad
G_{\rm{eff bgd}} = \frac{1}{1 + \ovf + \ovU } \;.
\ee

Finally, the growth equation governing the evolution of $\delta$ can be directly characterized by $G_{\rm{eff} \Psi} $,
\be
\ddot{\delta} + 2H\dot{\delta} 
= \frac{3}{2}H_0^2 \Omega_m a^{-3} (1+Y) \delta\;, \quad \mbox{with }  
1+ Y \equiv G_{\rm{eff} \Psi}\;. 
\eql{growth-eq-Y}
\ee
Alternatively, using the parameter $\mu$ defined in Ref. \cite{def-of-mu},
\be
\Psi \equiv (1+\mu)\Psi_{\rm GR}\;,
\eql{mu-def}
\ee
we have 
\be
\ddot{\delta} + 2H\dot{\delta}  
= \frac{3}{2}H_0^2 \Omega_m a^{-3} (1+ \mu) \delta\;, \quad \mbox{where }  
1+ \mu = (1-\eta) G_{\rm{eff} \Phi}\;. 
\eql{growth-eq-mu}
\ee
By comparing \Ec{growth-eq-Y} and \Ec{growth-eq-mu}, one can easily see that $Y=\mu$.
The right panel of Fig. \ref{fig:Geff-Y=mu}  
presents $Y = \mu$ as a function of redshift, which agrees with $Y$ in Ref. \cite{NCA-2017}.
The sign of $Y = \mu$ being negative means that the growth of $\delta$ gets suppressed in the DW model compared to the one in GR. 
Reference \cite{NCA-2017} has pointed out that this $G_{\rm{eff} \Psi}$ being negative is related to the violation of the ghost-free condition in the localized version  
\cite{ghost-nonlocal-Koivisto-0807, ghost-nonlocal-NOSZ-1010, Bamba-1104, ghost-nonlocal-ZS-1108,  ghost-Amendola-0409}. 
\footnote{It should be noted that Ref. \cite{NCA-2017} checked the violation of the ghost-free condition $6f'(X) > 1 + f(X) + U > 0 $ with only the background solutions $f(\ovX)$ and $\ovU$. It would be interesting to check the condition including linear perturbations, which would require solving the full linear perturbation equations.}
On the other hand, Refs. \cite{W-review-2014, DW-2013} have shown the absence of ghosts in the original nonlocal version. 
In any case, at the linear perturbation level, the nonlocal and localized versions give the same solutions for perturbation variables as long as the initial conditions are set the same, and no kinetic instability is observed (i.e., no linear perturbations blow up). 

\section{Discussion}
\label{discussion}

We have investigated the origin of two opposite results for the growth rate $f\sigma_8$ in the DW nonlocal gravity model.
The growth rate obtained by one group \cite{DP-2013} was higher than that of $\Lambda$CDM, but
the one obtained by the other group \cite{NCA-2017} was lower than $\Lambda$CDM.    

The evolution equations for the perturbation variables $\Phi, \,\Psi$, and $\delta$ derived by the two groups (the former in the original nonlocal formulation and the latter in a localized version) were equivalent, and the only difference was in the implementation of the sub-horizon limit on the two auxiliary equations for $\delta X$ and $\delta U$. 
In the sub-horizon scales of $k \gg Ha$, 
time derivatives are typically dropped;
hence, the limit is also called the quasi-static limit.
While the former group took the quasi-static limit only on the source terms of the rhs to keep the time derivatives of $\delta X$ and $\delta U$,
\begin{eqnarray*}
\Bigl(-\partial_t^2 - 3H \partial_t -\frac{k^2}{a^2} \Bigr) \delta X &\!\!\!=\!\!\!& S_{\delta X} 
\;, \\
\Bigl(-\partial_t^2 - 3H \partial_t -\frac{k^2}{a^2} \Bigr) \delta U &\!\!\!=\!\!\!& S_{\delta U}
\;, 
\end{eqnarray*}
the latter took the limit both on the lhs and rhs, which turned the differential equations into algebraic ones:
\begin{eqnarray*}
-\frac{k^2}{a^2} \delta X &\!\!\!=\!\!\!& S_{\delta X} \;,\\ 
-\frac{k^2}{a^2} \delta U &\!\!\!=\!\!\!& S_{\delta U} \;. 
\end{eqnarray*} 

The current author initially suspected that this different implementation of 
the quasi-static limit was the main source of the discrepancy. 
However, as one can see from their equations, 
the auxiliary fields are forced harmonic oscillators; henceforth, they eventually follow the applied force on the source term. Moreover, the transient oscillation of $\delta U$ in the beginning (at high redshifts) is erased by the factor of $\ovf'$ (which is almost zero at high redshifts $z \gtrsim 5$) in the equations of $\Phi$ and $\Psi$, and therefore the two implementations give almost the same solutions for $\Phi, \; \Psi$ and $\delta$. Working on the nonlocal form only, the author did not realize this now-simple behavior, and she made a mistake in the numerical calculation code, which unfortunately led to the result
opposite of the one by Ref. \cite{NCA-2017}. 

We summarize what should be corrected 
in our previous papers \cite{PD-2012, DP-2013, PS-2016}. 
In Ref. \cite{PD-2012}, all equations and the numerical results plotted in figures are actually correct.
However, it should be noted that the results of $G_{\rm{eff} \Phi}$ and $\eta$ were obtained by solving the perturbation equations perturbatively, which means 
by inserting the standard $\Lambda$CDM solution for $\Phi$ and $\Psi$ into the new terms. 
Since $\Phi$ and $\Psi$ in the DW model turn out to be quite different from $\Phi_{\rm GR}$ and $\Psi_{\rm GR}$, these are somewhat rough estimates. 
In Ref. \cite{DP-2013}, all equations are correct, but the numerical solutions plotted in the figures are wrong; therefore, the discussions based on those solutions should be discarded.
Reference \cite{PS-2016} attempted to lower the growth of $\delta$ by changing the background expansion rather than that of $\Lambda$CDM, motivated by the wrong conclusion of Ref. \cite{DP-2013} that {\it the DW model cannot do better than 
$\Lambda$CDM in suppressing growth when its background is fixed by $\Lambda$CDM}.   
All equations there are also correct, but the solutions for $\delta$ plotted in Figs. 2 and 3 are subject to change. The tendency that the equation of state $w$ less negative than $-1$ gives more suppressed growth might still remain because the dimensionless Hubble parameter being in the denominator in the source term of the growth equation ((17) in Ref. \cite{PS-2016}) was a key factor lowering the growth. However, more careful analysis is required to confirm it. Furthermore, it would be still worth it to check the growth of perturbations in the DW model with different background expansions rather than $\Lambda$CDM.    

After all, we extend the conclusion of Ref. \cite{NCA-2017} to include the equivalence of the nonlocal and localized formulations on the linear perturbation level so that both versions of the DW model make identical predictions: the lower growth of the matter overdensity $\delta$ compared to the case of $\Lambda$CDM. 
We also remark that the entire analysis on the perturbation equations in this paper was performed for only one scale $k=100H_0 = 0.03h \rm{Mpc}^{-1}$ and the quasi-static approximation worked 
very well in this sub-horizon scale. 
This  raises the question of in what range of $k$ the approximation holds. Solving the full linear perturbation equations
without taking the quasi-static limit would help to answer this question.
We note that the full linear equations for $\Phi$ and $\Psi$  are already given in  \Ec{full-linear-00-eq} and \Ec{full-linear-traceless-eq}, and for $\delta,  \delta X$, and $\delta U$, we only need to recover the time derivatives dropped in the source terms. 
It would be also interesting to check the dependence of the solutions on the initial conditions. 
In practice, the localized version is much simpler than the nonlocal one and is expected to be relatively easily incorporated with extensive cosmological simulation codes such as CAMB \cite{CAMB} and CLASS \cite{CLASS}. 
There are also theoretical issues like the apparent violation of the ghost-free condition in the localized formulation. However, we agree with the authors of Ref. \cite{NCA-2017} that the DW nonlocal model gives interesting predictions at linear perturbation level and deserves further consideration.  

\section*{Acknowledgements}

The author is grateful to Luca Amendola, Adrian Fernandez Cid, and Henrik Nersisyan for bringing the discrepancy to her attention and for the helpful discussions. The author also thanks Sibel Boran, Scott Dodelson, Emre Onur Kahya, St\'ephane Ili\'c, Arman Shafieloo, Constantinos Skordis, and Richard Paul Woodard for their insightful comments. 
The research leading to these results has received funding from the European Research Council under the European Union's Seventh Framework 
Programme (FP7/2007-2013) / ERC Grant No. 617656, ``Theories and Models of the Dark Sector: Dark Matter, Dark Energy and Gravity."

\bibliographystyle{unsrtnat}
\bibliography{references}

\begin{thebibliography}{50}
\providecommand{\natexlab}[1]{#1}
\providecommand{\url}[1]{\texttt{#1}}
\expandafter\ifx\csname urlstyle\endcsname\relax
  \providecommand{\doi}[1]{doi: #1}\else
  \providecommand{\doi}{doi: \begingroup \urlstyle{rm}\Url}\fi




\bibitem{DW-2007} S. Deser and R. P. Woodard, Phys. Rev. Lett. {\bf 99}, 111301 (2007), arXiv:0706.2151.

\bibitem{Joukovskaya-2007} 
L. Joukovskaya, Phys. Rev. D {\bf 76}, 105007 (2007), arXiv:0707.1545.

\bibitem{NO-2007}  
S. Nojiri and S. D. Odintsov Phys. Lett. B {\bf 659}, 821 (2008), arXiv:0708.0924.

\bibitem{CMN-2008} 
G. Calcagni, M. Montobbio and G. Nardelli, Phys. Lett. B {\bf 662}, 285 (2008), arXiv:0712.2237.

\bibitem{NO-2008} 
S. Jhingan, S. Nojiri, S. D. Odintsov, M. Sami, I. Thongkool and S. Zerbini, Phys. Lett. B {\bf 663}, 424 (2008), arXiv:0803.2613.

\bibitem{Koivisto-0803} 
T. S. Koivisto, Phys. Rev. D {\bf 77}, 123513 (2008), arXiv:0803.3399.

\bibitem{ghost-nonlocal-Koivisto-0807} 
T. S. Koivisto, Phys. Rev. D {\bf 78}, 123505 (2008),  arXiv:0807.3778. 

\bibitem{CENO-2009} 
S. Capozziello, E. Elizalde, S. Nojiri and S. D. Odintsov, Phys. Lett. B {\bf 671}, 193 (2009), arXiv:0809.1535.

\bibitem{Koshelev-2009} 
N. A. Koshelev, Grav. Cosmol. {\bf 15}, 220 (2009), arXiv:0809.4927.

\bibitem{DW-2009} C. Deffayet and R. P. Woodard, J. Cosmol. Astropart. Phys. 08 (2009) 023, arXiv:0904.0961.

\bibitem{Biswas-2010} 
T. Biswas, T. S. Koivisto and A. Mazumdar, J. Cosmol. Astropart. Phys. 11 (2010) 008, arXiv:1005.0590.

\bibitem{ghost-nonlocal-NOSZ-1010} 
S. Nojiri, S. D. Odintsov, M. Sasaki and Y.-l. Zhang, Phys. Lett. B {\bf 696}, 278 (2011), arXiv:1010.5375. 


\bibitem{Bamba-1104}
K. Bamba, S. Nojiri, S. D. Odintsov, and M. Sasaki, Gen. Rel. Grav. {\bf 44}, 1321 (2012), arXiv:1104.2692.


\bibitem{Barvinsky-1107} 
A. O. Barvinsky, Phys. Lett. B {\bf 710}, 12 (2012), arXiv:1107.1463. 

\bibitem{ghost-nonlocal-ZS-1108} 
Y.-l. Zhang and M. Sasaki, Int. J. Mod. Phys. D {\bf 21},  1250006 (2012),  arXiv:1108.2112.

\bibitem{EPV-1110} 
E. Elizalde, E. O. Pozdeeva and S. Yu. Vernov, Phys. Rev. D {\bf 85}, 044002 (2012), arXiv:1110.5806.


\bibitem{Barvinsky-1112} 
A.O. Barvinsky, Phys. Rev. D {\bf 85}, 104018 (2012), arXiv:1112.4340.


\bibitem{PD-2012} S. Park and S. Dodelson, Phys. Rev. D {\bf 87}, 024003 (2013), arXiv:1209.0836.

\bibitem{Barvinsky-1209} 
A. O. Barvinsky and Y. V. Gusev, Phys. Part. Nucl. {\bf 44}, 213 (2013), arXiv:1209.3062.

\bibitem{EPV-1209} 
E. Elizalde, E. O. Pozdeeva and S. Yu. Vernov, Class. Quant. Grav. {\bf 30},  035002 (2013), arXiv:1209.5957.

\bibitem{EPVZ-1302} 
E. Elizalde, E. O. Pozdeeva, S. Yu. Vernov and Y.-l. Zhang, J. Cosmol. Astropart. Phys. 07 (2013) 034, arXiv:1302.4330.

\bibitem{Maggiore-1307} 
M. Maggiore, Phys. Rev. D {\bf 89},  043008 (2014), arXiv:1307.3898.

\bibitem{DW-2013} S. Deser  and R. P. Woodard, J. Cosmol. Astropart. Phys. 11 (2013) 036, arXiv:1307.6639.

\bibitem{DP-2013} S. Dodelson and S. Park,  Phys. Rev. D {\bf 90}, 043535 (2014), arXiv:1310.4329.

\bibitem{Maggiore-1311.3421} 
S. Foffa, M. Maggiore, and E. Mitsou, Phys. Lett. B {\bf 733}, 76 (2014), arXiv:1311.3421.

\bibitem{Maggiore-1311.3435} 
S. Foffa, M. Maggiore, and E. Mitsou, Int. J. Mod. Phys. A {\bf 29}, 1450116 (2014), arXiv:1311.3435.

\bibitem{W-review-2014}   R. P. Woodard, Found. Phys. {\bf 44}, 213 (2014), arXiv:1401.0254.

\bibitem{Maggiore-1401} 
A. Kehagias and M. Maggiore, J. High Energy Phys. 08 (2014) 029, arXiv:1401.8289.

\bibitem{MM-2014} M. Maggiore and M. Mancarella, Phys. Rev. D {\bf 90}, 023005 (2014), arXiv: 1402.0448.

\bibitem{Maggiore-1403} 
Y. Dirian, S. Foffa, N. Khosravi, M. Kunz, and M. Maggiore, J. Cosmol. Astropart. Phys. 06 (2014) 033, arXiv:1403.6068.

\bibitem{Koivisto-1406} 
A. Conroy, T. Koivisto, A. Mazumdar, and A. Teimouri, Class. Quant. Grav. {\bf 32}, 015024 (2015),  arXiv:1406.4998.

\bibitem{BLHBP-1408} 
A. Barreira, B. Li, W. A. Hellwing, C. M. Baugh, and S. Pascoli, J. Cosmol. Astropart. Phys. 09 (2014) 031, arXiv:1408.1084.

\bibitem{DM-1408} 
Y. Dirian and E. Mitsou, JCAP 1410 (2014) 065, arXiv:1408.5058.

\bibitem{Maggiore-1411} 
Y. Dirian, S. Foffa, M. Kunz, M. Maggiore, and V. Pettorino, J. Cosmol. Astropart. Phys. 04 (2015) 044, arXiv:1411.7692.

\bibitem{ZWLLCCS-1511}
X. Zhang, Y. Wu, S. Li, Y. Liu, B. Chen, Y. Chai, and S. Shu, J. Cosmol. Astropart. Phys. 07 (2016) 003, arXiv:1511.05238. 

\bibitem{Maggiore-1512} 
G. Cusin, S. Foffa, M. Maggiore, and M. Mancarella, Phys. Rev. D {\bf 93} (2016) 043006, arXiv:1512.06373.


\bibitem{ZKSZ-1601}
Y.-l. Zhang, K. Koyama, M. Sasaki, and G. Zhao, J. High Energy Phys. 03 (2016) 039, arXiv:1601.03808.

\bibitem{Maggiore-1602.01078} 
G. Cusin, S. Foffa, M. Maggiore, and M. Mancarella, Phys. Rev. D {\bf 93}, 083008 (2016), arXiv:1602.01078.

\bibitem{Maggiore-1602.03558} 
Y. Dirian, S. Foffa, M. Kunz, M. Maggiore, and V. Pettorino, J. Cosmol. Astropart. Phys. 05 (2016) 068,  arXiv:1602.03558.

\bibitem{Maggiore-1603} 
M. Maggiore, Phys. Rev. D {\bf 93},  063008 (2016), arXiv:1603.01515.

\bibitem{NAAKR-1606} 
H. Nersisyan, Y. Akrami, L. Amendola, T. S. Koivisto, and J. Rubio, Phys. Rev. D {\bf 94},  043531 (2016), arXiv:1606.04349.

\bibitem{Maggiore-review} 
M. Maggiore, Fundam. Theor. Phys. {\bf 187}, 221 (2017), arXiv:1606.08784.

\bibitem{PS-2016} S. Park and A. Shafieloo, Phys. Rev. D {\bf 95}, 064061 (2017), arXiv:1608.02541.

\bibitem{WZWZC-1611} 		
Y. Wu, X. Zhang, M. Wu, N. Zhang, and B. Chen, Chin. Phys. Lett. {\bf 34}, 079801 (2017),  arXiv:1611.07209.

\bibitem{NCA-2017} H. Nersisyan, A. F. Cid, and L. Amendola, J. Cosmol. Astropart. Phys. 04 (2017) 046, arXiv:1701.00434. 


\bibitem{DDSKR-1701}
I. Dimitrijevic, B. Dragovich, J. Stankovic, A. S. Koshelev, and Z. Rakic, Springer Proc. Math. Stat. {\bf 191},  35 (2016), arXiv:1701.02090.

\bibitem{VAAS-2017} V. Vardanyan, Y. Akrami, L. Amendola, and A. Silvestri, arXiv:1702.08908. 

\bibitem{Dirian-1704}
Y. Dirian,  Phys. Rev. D {\bf 96}, 083513 (2017), arXiv:1704.04075.

\bibitem{ABN-1707} 	 	  
L. Amendola,  N. Burzilla, and H. Nersisyan, Phys. Rev. D {\bf 96}, 084031 (2017), arXiv:1707.04628. 

\bibitem{Roobiat-2017} 	  
K.Y. Roobiat and R. Pazhouhesh,  Int. J. Mod. Phys. D {\bf 26}, 1750134 (2017). 

\bibitem{ZWYZCZ-2017} 	
X. Zhang, Y. Wu, W. Yang, C. Zhang, B. Chen, and N. Zhang, Sci. China Phys. Mech. Astron. {\bf 60}, 100411 (2017).

\bibitem{Sebastian-1708}
S. Bahamonde, S. Capozziello, and K. F. Dialektopoulos, Eur. Phys. J. C {\bf 77}, 722 (2017), arXiv:1708.06310. 

\bibitem{Sebastian-1709}
S. Bahamonde, S. Capozziello, M. Faizal, and R. C. Nunes, Eur. Phys. J. C {\bf 77}, 628 (2017), arXiv:1709.02692. 

\bibitem{Bamba-1711}
K. Bamba, D. Momeni, and M. A. Ajmi, arXiv:1711.10475.




\bibitem{Parker-1985} 
L. Parker and D. J. Toms, Phys. Rev. D {\bf 32}, 1409 (1985). 

\bibitem{Banks-1988} 
T. Banks, Nucl. Phys. B {\bf 309}, 493 (1988). 

\bibitem{Wetterich-1998} 
C. Wetterich, Gen. Rel. Grav. {\bf 30}, 159 (1998), gr-qc/9704052.

\bibitem{Barvinsky-2003} 
A. O. Barvinsky, Phys. Lett. B {\bf 572}, 109 (2003), hep-th/0304229.

\bibitem{Hamber-2005} 
H. W. Hamber and R. M. Williams, Phys. Rev. D {\bf 72}, 044026 (2005), hep-th/0507017.

\bibitem{Biswas-2006} 
T. Biswas, A. Mazumdar, and W. Siegel, J. Cosmol. Astropart. Phys. 03 (2006) 009, hep-th/0508194.

\bibitem{Mazzitelli-2007} 
D. Lopez Nacir and F. D. Mazzitelli, Phys. Rev. D {\bf 75},  024003 (2007), hep-th/0610031.

\bibitem{Ferreira-2013} 
P. G. Ferreira and A. L. Maroto, Phys. Rev. D {\bf 88}, 123502 (2013), arXiv:1310.1238.

\bibitem{Donoghue-2014} 
J. F. Donoghue and B. K. El-Menoufi, Phys. Rev. D {\bf 89}, 104062 (2014), arXiv:1402.3252.

\bibitem{Calmet-2015} 
X. Calmet, D. Croon, and C. Fritz, Eur. Phys. J. C {\bf 75}, 605 (2015), arXiv:1505.04517.

\bibitem{Donoghue-2015} 
J. F. Donoghue and B. K. El-Menoufi, J. High Energy Phys. 10 (2015) 044, arXiv:1507.06321.


\bibitem{Oda-1704}
I. Oda, Phys. Rev. D {\bf 95}, 104020 (2017), arXiv:1704.05619.

\bibitem{Oda-1706}
I. Oda, Phys. Rev. D {\bf 96}, 024027 (2017), arXiv:1706.01683.

\bibitem{Oda-1709}
I. Oda, arXiv:1709.08189.


\bibitem{Mazzitelli-2017} 	
M. Elias, F. D. Mazzitelli, and L.G. Trombetta, Phys. Rev. D {\bf 96}, 105007 (2017), arXiv:1709.10435.

\bibitem{Bautista-2017} 	
T. Bautista, A. Benevides, and A. Dabholkar, arXiv:1711.00135. 

\bibitem{Calmet-2017} 		
S. O. Alexeyev, X. Calmet, and B. N. Latosh, Phys. Lett. B {\bf 776}, 111 (2018), arXiv:1711.06085.

\bibitem{Oda-1711}
I. Oda, arXiv:1711.10476.








\bibitem{TW-0904} 
N. C. Tsamis and R. P. Woodard,  Phys. Rev. D {\bf 80},  083512 (2009), arXiv:0904.2368.

\bibitem{TW-1006} 
N. C. Tsamis and R. P. Woodard,  Phys. Rev. D {\bf 82}, 063502 (2010), arXiv:1006.4834.

\bibitem{TW-1405} 
N. C. Tsamis and R. P. Woodard,  J. Cosmol. Astropart. Phys. 09 (2014) 008, arXiv:1405.4470.

\bibitem{KKM-1604} 
A. S. Koshelev, K. S. Kumar, and P. V. Moniz, Phys. Rev. D {\bf 96}, 103503 (2017), arXiv:1604.01440.

\bibitem{TW-1606} 
N. C. Tsamis and R. P. Woodard,  Phys. Rev. D {\bf 94}, 043508 (2016), arXiv:1606.06967.




\bibitem{SW-2003} 
M. E. Soussa and R. P. Woodard, Class. Quant. Grav. {\bf 20}, 2737 (2003), astro-ph/0302030.

\bibitem{DEW-1106} 
C. Deffayet, G. Esposito-Farese, and R. P. Woodard, Phys. Rev. D {\bf 84}, 124054 (2011), arXiv:1106.4984.

\bibitem{W-MOND-review-1403} 
R. P. Woodard, Can. J. Phys. {\bf 93}, 242 (2015),  arXiv:1403.6763.

\bibitem{DEW-1405} 
C. Deffayet, G. Esposito-Farese, and R. P. Woodard,  Phys. Rev. D {\bf 90}, 064038 (2014), 
Addendum: Phys.Rev. D {\bf 90}, 089901 (2014), arXiv:1405.0393 

\bibitem{KRSTWX-1608} 
M. Kim, M.H. Rahat, M. Sayeb, L. Tan, R. P. Woodard, and B. Xu, Phys. Rev. D {\bf 94}, 104009 (2016), arXiv:1608.07858. 


\bibitem{Bertschinger-2006}
E. Bertschinger, Astrophys. J. {\bf 648}, 797 (2006), astro-ph/0604485.

\bibitem{Hu-Sawicki-0708}
W. Hu and I. Sawicki, Phys. Rev. D {\bf 76}, 104043 (2007), arXiv:0708.1190.

\bibitem{Baker:2012zs} 
T. Baker, P. G. Ferreira, and C. Skordis, Phys. Rev. D {\bf 87}, 024015 (2013), arXiv:1209.2117.

\bibitem{Huterer:2013xky} 
D. Huterer, D. Kirkby, R. Bean, A. Connolly, K. Dawson, et al.,  Astropart. Phys. {\bf 63}, 23 (2015), arXiv:1309.5385.


\bibitem{WMAP-5year} J. Dunkley et al. [WMAP Collaboration], Astrophys. J. Suppl. {\bf 180}, 306 (2009), arXiv:0803.0586. E. Komatsu et al. [WMAP Collaboration], Astrophys. J. Suppl. {\bf 180}, 330 (2009), arXiv:0803.0547.

\bibitem{Planck2015-XIII} Planck Collaboration, Astron. Astrophys. {\bf 594}, A13 (2015), arXiv:1502.01589.

\bibitem{Ma-Bert1995} C.-P. Ma and E. Bertschinger, Astrophys. J. {\bf 455}, 7 (1995), arXiv:astro-ph/9506072.

\bibitem{Dodelson-book} S. Dodelson, {\it Modern Cosmology} (Academic Press, San Diego, 2003).

\bibitem{Amendola-Tsujikawa-book} L. Amendola and S. Tsujikawa, {\it Dark Energy: Theory and Observations} (Cambridge University Press, 2010).

\bibitem{def-of-mu} F. Simpson, C. Heymans, D. Parkinson, C. Blake, M. Kilbinger, et al., Mon. Not. Roy. Astron. Soc. \textbf{429}, 2249 (2013), arXiv:1212.3339. 

\bibitem{ghost-Amendola-0409} 
L. Amendola, Phys. Rev. Lett. {\bf 93}, 181102 (2004),  hep-th/0409224.

\bibitem{CAMB}	
A. Lewis and A. Challinor, {\it CAMB: Code for Anisotropies in the Microwave Background}, Astrophysics Source Code Library (2011), ascl:1102.026.

\bibitem{CLASS}
D. Blas, J. Lesgourgues, and T. Tram, JCAP 07 (2011) 034, arXiv:1104.2933.

\end{thebibliography}

\end{document}